\documentclass[11pt,letterpaper]{article}
\pdfoutput=1
\usepackage{jcappub}
\usepackage{color}
\usepackage{graphicx}
\usepackage{tabularx}
\usepackage{xspace}
\usepackage{caption}
\usepackage{subcaption}
\usepackage{float}

\usepackage{verbatim}
\usepackage{amsmath}
\usepackage{amssymb}
\usepackage{url}
\usepackage{bbold}
\usepackage{slashed}
\usepackage{array}

\usepackage{multirow}
\usepackage{threeparttable}
\usepackage{paralist}
\usepackage{xspace}
\usepackage{upgreek}
\usepackage{lipsum}

\newcommand\colvec[3][]{\begin{pmatrix}\ifx\relax#1\relax\else#1\\\fi#2\\#3\end{pmatrix}}

\definecolor{darkmagenta}{rgb}{0.55, 0.0, 0.55}

\newcommand{\beq}{\begin{equation}}
\newcommand{\beqn}{\begin{eqnarray}}
\newcommand{\eeq}{\end{equation}}
\newcommand{\eeqn}{\end{eqnarray}}
\newcommand\numberthis{\addtocounter{equation}{1}\tag{\theequation}}
\newcommand{\dbar}{\ensuremath{\mathchar'26\mkern-12mu d}}
\newcommand\order[1]{{\cal O}#1}

\DeclareRobustCommand{\Eq}[1]{Eq.~(\ref{#1})}

\DeclareRobustCommand{\Sec}[1]{Sec.~\ref{#1}}

\DeclareRobustCommand{\Fig}[1]{Fig.~\ref{#1}}

\renewcommand{\vec}[1]{\mathbf{#1}}

\begin{document}

\title{The Trispectrum in the Effective Field Theory of Large Scale Structure}
\author{Daniele Bertolini,}
\author{Katelin Schutz,}
\author{Mikhail P. Solon,}
\author{and Kathryn M. Zurek}

\affiliation{Berkeley Center for Theoretical Physics, University of California, Berkeley, CA 94720 }
\affiliation{Theoretical Physics Group, Lawrence Berkeley National Laboratory, Berkeley, CA 94720}

\emailAdd{dbertolini@lbl.gov}
\emailAdd{kschutz@berkeley.edu}
\emailAdd{mpsolon@lbl.gov}
\emailAdd{kmzurek@lbl.gov}

\date{\today}

\abstract{We compute the connected four point correlation function (the trispectrum in Fourier space) of cosmological density perturbations at one-loop order in Standard Perturbation Theory (SPT) and the Effective Field Theory of Large Scale Structure (EFT of LSS). This paper is a companion to our earlier work on the non-Gaussian covariance of the matter power spectrum, which corresponds to a particular wavenumber configuration of the trispectrum.   In the present calculation, we highlight and clarify some of the subtle aspects of the EFT framework that arise at third order in perturbation theory  for general wavenumber configurations of the trispectrum. We consistently incorporate vorticity and non-locality in time into the EFT counterterms and lay out a complete basis of building blocks for the stress tensor. We show predictions for the one-loop SPT trispectrum and the EFT contributions, focusing on configurations which have particular relevance for using LSS to constrain primordial non-Gaussianity.}

\maketitle

\section{Introduction}
\label{sec:intro}
Understanding large scale structure (LSS) provides myriad insights into both the physics of how density perturbations evolve as well as the physics governing the initial conditions set by inflation. Beyond the two-dimensional cosmic microwave background (CMB), which has already greatly constrained cosmological parameters and inflation, LSS can provide an even greater wealth of information owing to the simple fact that there are more samples of a given Fourier mode in a three-dimensional volume than in a two-dimensional slice. However, in order to extract this additional information, one must pay a price: perturbations in the CMB are governed by linear theory while understanding LSS requires that nonlinearities due to the formation of structure be taken into account. Most notably, perturbations become nonlinear as structures gravitationally collapse, inducing mode-coupling which becomes more and more substantial over time and at small scales. Thus, even if the initial density is a simple Gaussian random field, the evolution of perturbations renders the density field increasingly non-Gaussian. 

All of the information of a statistically homogeneous Gaussian random field is contained in the two point correlation function, since all higher order moments are given by Wick expansion. In Fourier space, density perturbations $\delta(\vec{k})$ that are Gaussian are thus characterized entirely by their power spectrum, defined as
\beq \langle \delta(\vec{k}_1) \delta(\vec{k}_2)\rangle = (2\pi)^3 \delta_D(\vec{k}_1+\vec{k}_2) P(k_1), \eeq where $\delta_D$ denotes the Dirac delta function, and the power spectrum is only a function of the magnitude $k_1=|\vec{k}_1|$.
However, since the density field becomes non-Gaussian at late times and small scales, higher order moments are rich in additional information. Beyond the two point function, the three point function measures the skewness, while the connected four point function measures kurtosis. In Fourier space, these correspond to the matter bispectrum and trispectrum, defined as \beq
\langle \delta(\vec{k}_1) \delta(\vec{k}_2) \delta(\vec{k}_3) ) \rangle = (2\pi)^3 \delta_D(\vec{k}_1+\vec{k}_2+\vec{k}_3) B(\vec{k}_1, \vec{k}_2, \vec{k}_3),
\eeq
\beq
\langle \delta(\vec{k}_1) \delta(\vec{k}_2) \delta(\vec{k}_3) \delta(\vec{k}_4) \rangle_c = (2\pi)^3 \delta_D(\vec{k}_1+\vec{k}_2+\vec{k}_3+\vec{k}_4) T(\vec{k}_1, \vec{k}_2, \vec{k}_3, \vec{k}_4)\,,
\eeq
where the subscript $c$ denotes the connected part of the correlation function.
Determining the origin of these higher order moments in LSS, whether primordial or from nonlinear structure formation, is of central importance for constraining the physics governing inflation ({\em e.g.}~\cite{1997PhRvD..56..535L,2002PhLB..524....5L,2002PhRvD..65j3505B,2003PhRvD..67b3503L}) via signatures of nonlinear clustering \cite{2002PhRvD..66f3008O,2006JCAP...05..004C,2010CQGra..27l4011D}.
In particular, it has been shown that a measurement of the trispectrum (which has weaker dependence on nonlinear clustering) may provide constraints on primordial non-Gaussianity that are complementary to those obtained from bispectrum measurements~\cite{2001ApJ...553...14V}.

With this as motivation, the purpose of this paper is to calculate the trispectrum in the weakly nonlinear regime. This work is a companion to our previous work on the covariance of the matter power spectrum, which corresponds to a particular wavenumber configuration of the trispectrum \cite{2015arXiv151207630B}. Making a theoretical prediction for LSS observables in the weakly non-linear regime is analytically challenging; even the largest scales are not safe from the effects of mode-coupling with small nonlinear scales. This mode coupling makes Standard Perturbation Theory (SPT) sensitive to ultraviolet (UV) modes in an unphysical way, and the Effective Field Theory of Large Scale Structure (EFT of LSS) has emerged as a useful tool for correcting this through systematically incorporating the feedback of small-scale nonlinearities on larger scales. This procedure is analogous to renormalization and allows us to extend our theoretical understanding of LSS down to smaller scales~\cite{2012JCAP...07..051B,2014PhRvD..89d3521H}. This extension is especially useful because there are far more samples of smaller scale modes in a given volume, which enhances their information content.

Already the EFT of LSS has been used to make predictions for various physical observables, such as the matter power spectrum~\cite{2012JHEP...09..082C} and the matter bispectrum, both with~\cite{2015JCAP...11..024A} and without~\cite{2014arXiv1406.4143A,2015JCAP...05..007B} primordial non-Gaussianity. For the one-loop power spectrum and bispectrum, renormalization of the leading UV sensitivity requires EFT counterterms at leading order (LO) and next-to-leading order (NLO) in powers of the linear density perturbation $\delta_1$, respectively.

In this paper, we make predictions in both SPT and the EFT of LSS for the trispectrum.  We present the first complete calculation of the four-point density correlation function in SPT at one-loop order for a general wavenumber configuration. In carrying out a renormalization of the leading UV sensitivity of the one-loop trispectrum, we require the fully general form of the EFT of LSS counterterms at next-to-next-to-leading order (NNLO), {\em i.e.}, through $\order(\delta_1^3)$. 

Working at NNLO, and with general wavenumber configurations, requires a thorough analysis of the unique technical features of the EFT of LSS, such as non-locality in time and the vorticity induced by effective operators. In particular, we show by direct calculation 
that the Eulerian formulation of non-locality in time is captured by counterterms that are local both in space and time, even at NNLO. Additionally, an essential ingredient of renormalization that first arises at this order is the vorticity induced by the stress tensor and heat conduction terms. We provide the necessary solution for the induced vorticity field, and illustrate its role in the context of a velocity field redefinition commonly employed in the literature. We provide the full set of EFT of LSS kernels relevant for the trispectrum, in a general but minimal form that can be applied for the calculation of the two-loop power spectrum, the covariance, and other observables at higher order. As an explicit application, we demonstrate the consistent renormalization of the power spectrum, bispectrum and trispectrum using our kernels. Once we have shown the consistency of our analytic results, we plot predictions for both the SPT and EFT contributions to the trispectrum in wavenumber configurations where primordial non-Gaussianity (both local and equilateral) is expected to be relatively large.

The rest of the paper is organized as follows. In Sec.~\ref{sec:EOM}, we review the SPT equations of motion and apply the EFT smoothing procedure to them. We correct previous expressions for the mode coupling functions involving vorticity, which become relevant at NNLO. 
The focus of Sec.~\ref{sec:stress} is to clarify issues that arise in constructing the EFT stress tensor at NNLO, such as the impact of time non-locality. We then construct the stress tensor and compute the fully general kernels for the trispectrum in the EFT of LSS. In Sec.~\ref{sec:trispectrum}, we demonstrate consistent renormalization of the SPT trispectrum. We also show both the SPT prediction and the EFT of LSS contributions for particular wevenumber configurations, which we have computed using {\tt FnFast}, our publicly available code for numerically computing SPT and EFT diagrams \cite{2015arXiv151207630B}. Concluding remarks follow in Sec.~\ref{sec:conclusions}. 

\section{Formalism}
\label{sec:EOM}
For completeness and to set our notation, this section provides a review of the equations of motion both in SPT and in the EFT of LSS, along with their perturbative solutions. Since much of this formalism is outlined broadly in the EFT of LSS literature, well-versed readers can skip directly to Eq.~\eqref{eq:couplings} where we correct previous results for the mode coupling functions involving vorticity, and to \Sec{sec:heat} where we illustrate the correspondence between heat conduction terms and the vorticity. These subtleties involving the vorticity were not relevant for the lower-order counterterms employed in previous EFT of LSS calculations, but are essential for the structure of EFT of LSS counterterms at NNLO.

\subsection{SPT}
To derive the Eulerian-space equations for a self-gravitating fluid of cold dark matter particles, we begin by considering the collisionless Boltzmann equation
\begin{align*}
\frac{d f}{dt} 
& = \frac{1}{a(\tau)}\frac{\partial f}{\partial \tau} +\frac{p^i}{a(\tau)^2m} \frac{\partial f}{\partial x^i}-m \frac{\partial f}{\partial p^i} \frac{\partial \phi}{\partial x_i} = 0, \numberthis\label{eq:Boltz}
\end{align*}
where $f$ is the phase space density, $m$ and $p^i$ are respectively the mass and momentum of the dark matter particles, and $a(\tau)$ is the scale factor. We have chosen to work with comoving coordinates and with conformal time $\tau$, defined relative to the comoving observer's time coordinate through $dt = a(t) d\tau$.
For our scales of interest, which are much less than $c/H$, we are justified in approximating the gravitational dynamics as  Newtonian. Thus, we have inserted $\phi$ which defines the gravitational potential in conformal Newtonian gauge.

Now, we define the first three moments of the phase space distribution function as 
\begin{align} 
\rho(\tau, \vec{x}) &= m \int d^3 p f(\tau, \vec{x}, \vec{p}), \\ 
\pi^i(\tau, \vec{x}) &=  \int d^3 p f(\tau, \vec{x}, \vec{p}) p^i, \\
\sigma^{ij} (\tau, \vec{x}) & = \frac{1}{m} \int d^3 p f(\tau, \vec{x}, \vec{p}) p^i p^j - \frac{\pi^i \pi^j}{\rho},
 \end{align}
which are the comoving mass and momentum densities and the comoving velocity dispersion tensor, respectively. In SPT, where the system is assumed to behave as a pressureless perfect fluid ({\em i.e.}~particles move in a single coherent flow), the velocity dispersion tensor and all the higher order moments are set to zero.  As we will show in the next section, the EFT of LSS includes a velocity dispersion tensor which parametrizes the effect of short-scale dynamics on long scales, including the effects of stream crossing. 
 
Setting $\sigma^{ij} (\tau, \vec{x})=0$ and taking the first two moments of the Boltzmann equation, we obtain 
\begin{align*}
\partial_\tau \rho + \frac{1}{a} \partial_i \pi^i &= 0, \numberthis\\
\partial_\tau \pi^i +\frac{1}{a} \partial_j \left(\frac{\pi^i \pi^j}{\rho}\right)+ a \rho\, \partial^i \phi &= 0. \numberthis \label{pieqn}
\end{align*}
It is more convenient to work in terms of the physical peculiar velocity $v^i$ which is related to the comoving momentum density via $\pi^i = \rho\, v^i a$. We also define density perturbations relative to the mean comoving density $\bar{\rho}$ as $\delta = \rho/\bar{\rho} - 1$. The perturbed continuity and Euler equations then become
\begin{align*} 
\partial_\tau \delta +  \partial_i (v^i (1+\delta)) &= 0,\numberthis\label{eq:Cont}\\
\partial_\tau v^i +  v^j \partial_j v^i + v^i \mathcal{H} +  \partial^i \phi& = 0, \numberthis\label{eq:Euler}
\end{align*}
where $\mathcal{H}$ is the conformal Hubble parameter.
We can decompose the velocity into its divergence $\theta = \partial_i v^i$ and its divergenceless curl ({\em i.e.} vorticity) $\omega^i = \epsilon^{ijk} \partial_j v_k$ as
\beq 
v^i = \frac{\partial^i}{\partial^2} \theta - \epsilon^{ijk} \frac{\partial_j}{\partial^2} \omega_k ,
\eeq
where $\epsilon^{ijk}$ is the Levi-Civita tensor. We can now substitute this decomposition of the velocity into Eqs.~\eqref{eq:Cont} and \eqref{eq:Euler} and take the Fourier transform. The continuity and Euler equations become
\begin{align*}
\partial_\tau \delta(\vec{k}) +  \theta(\vec{k}) &= - \int d^3q \Big( \alpha (\vec{q}, \vec{k}-\vec{q}) \theta(\vec{q})  - \alpha_\omega^i (\vec{q}, \vec{k}-\vec{q})\omega_i(\vec{q}) \Big) \delta(\vec{k}- \vec{q}), \numberthis\label{deltaeqn}\\
\partial_\tau \theta(\vec{k})+  \mathcal{H} \theta(\vec{k})+  \frac{3}{2} \mathcal{H}^2 \Omega_m\delta(\vec{k}) &=-\int d^3 q \bigg( \beta(\vec{q}, \vec{k}-\vec{q}) \theta(\vec{q}) \theta(\vec{k} - \vec{q})\\
-\vec{\beta}_\omega^i(\vec{q}, \vec{k}&-\vec{q})  \vec{\omega}_i(\vec{q})\theta(\vec{k} - \vec{q}) + \beta_{\omega\omega}^{ij}(\vec{q}, \vec{k}-\vec{q}) \,\omega_i (\vec{q})\,\omega_j( \vec{k}-\vec{q})\bigg), \numberthis\label{thetaeqn}\\
\partial_\tau \omega^i(\vec{k}) +\mathcal{H} \omega^i(\vec{k}) &= -\int d^3 q  \bigg(-\gamma^{ij}_\omega(\vec{q}, \vec{k}-\vec{q}) \omega_j(\vec{q}) \theta(\vec{k}-\vec{q})\\
&\quad\quad\quad\quad\quad\quad\quad +\gamma^{ijk}_{\omega\omega}(\vec{q}, \vec{k}-\vec{q}) \omega_j(\vec{q}) \omega_k(\vec{k}-\vec{q}) \bigg)\,, \numberthis \label{omegaeq}
\end{align*}
where we have used the Poisson equation for the Newtonian potential in comoving coordinates, 
\beq 
\partial^2 \phi = \frac{3}{2} \mathcal{H}^2 \Omega_m \delta,
\eeq
where $\Omega_m$ is the matter density.
The mode-coupling functions above are defined as 
\begin{align}\label{eq:couplings}
\alpha(\vec{k}_1, \vec{k}_2) &= \frac{\vec{k}_1 \cdot \vec{k}}{k_1^2},\\
\alpha_\omega^i(\vec{k}_1, \vec{k}_2)& = \frac{(\vec{k}_2 \times \vec{k}_1)^i}{k_1^2},\\
 \beta(\vec{k}_1, \vec{k}_2)& = \frac{k^2 (\vec{k}_1\cdot \vec{k}_2)}{2\, k_1^2 k_2^2},\\
 \beta_\omega^i(\vec{k}_1, \vec{k}_2) &= \frac{(2 (\vec{k}_1\cdot\vec{k}_2) + k_2^2)(\vec{k}_2\times \vec{k}_1)^i}{k_1^2 k_2^2},\\
 \beta_{\omega\omega}^{ij}(\vec{k}_1, \vec{k}_2) &= \frac{(\vec{k}_2\times \vec{k}_1)^i (\vec{k}_1\times \vec{k}_2)^j}{k_1^2 k_2^2},\\
\gamma^{ij}_\omega (\vec{k}_1, \vec{k}_2) & =\frac{k_2^ik^j-(\vec{k}\cdot\vec{k}_2)\delta^{ij}}{k_2^2},\\
\gamma_{\omega \omega}^{ijk}(\vec{k}_1, \vec{k}_2) &=  { \epsilon^{imj}k_m k_1^k-(\vec{k} \times \vec{k}_1)^i \delta^{jk} \over k_1^2} \,,
\end{align}
where $\vec{k} = \vec{k}_1 + \vec{k}_2$.\footnote{Note the useful identity $\epsilon_{ijk} \partial^j (v \cdot \partial v^k ) = - \epsilon_{ijk} \partial^j (\epsilon^{kst}v_s \omega_t)$.}
We note that the kernels $\beta_\omega^i$ and $\gamma_{\omega \omega}^{ijk}$ above differ from those appearing in Refs.~\cite{2015arXiv150907886A} and \cite{Pueblas:2008uv}. In particular, the correct $\beta_\omega^i$ kernel is crucial for a consistent renormalization of the trispectrum. We will discuss the role of vorticity for the trispectrum calculation in more detail in \Sec{sec:heat}.

For convenience of solving the equations of motion, we diagonalize the left hand side of the equations. To make this compact, we define $S_\alpha$ and $S_\beta$ to represent the entire RHS of Eqs.~\eqref{deltaeqn} and \eqref{thetaeqn}. Noting that $\partial_\tau = \mathcal{H} a \partial_a$ and that $\mathcal{H} = \mathcal{H}_0/\sqrt{a}$, and working in the Einstein De Sitter (EDS) case of $\Omega_m=1$, the SPT equations of motion are
 \begin{align} \label{diagdelta}
\mathcal{H}^2 \left(-a^2 \partial_a^2 - \frac{3}{2} a\partial_a \delta + \frac{3}{2}\right) \delta &= S_\beta-\mathcal{H}\partial_a (a S_\alpha),\\
 \mathcal{H} \left(1 - \frac{5}{2} a \partial_a - a^2 \partial_a^2   \right) \theta &= \frac{3}{2}\mathcal{H} S_\alpha-  \partial_a(a S_\beta ).
 \label{diagtheta}\end{align}
Once we establish a perturbative ansatz, we will be able to algebraically solve these differential equations for $\delta$ and $\theta$ order by order. 

\subsection{EFT of LSS}

In SPT, one solves Eqs.~(\ref{deltaeqn}-\ref{omegaeq}) perturbatively. Even if one focuses on large scales only, beyond tree level, the perturbative solution involves integrals over short-scale modes. In this large-wavenumber regime, not only is perturbation theory invalid, but the perfect fluid description itself breaks down (see, {\em e.g.}, \cite{2015arXiv150207389M}; in one dimension the perturbative series can be resummed, but it does not accurately reproduce simulations), as it does not describe stream crossing effects, which become relevant at small scales.

In the EFT of LSS one derives equations for smoothed long-wavelength modes only, where the feedback of the short-wavelength modes is parametrized in terms of effective corrections to the continuity and Euler equations. These corrections can be systematically organized in powers of the smoothed fields and derivatives. Given a generic field $\varphi(\vec{x})$, we define the long-wavelength part $\varphi_l(\vec{x})$ by convolving with a window function $W_\Lambda$ which averages over scales smaller than the characteristic scale $1/\Lambda$,
\beq \label{eq:EFTwindow}
\varphi_l (\vec{x}) = \int d^3x'  \,W_\Lambda (\vec{x} - \vec{x}')\, \varphi(\vec{x}').
\eeq 
One can take the smoothing function to be a Gaussian $W_\Lambda\propto\exp\left(-\frac{1}{2}|\vec{x} - \vec{x}'|^2\Lambda^2\right)$, for example. We can then follow the same steps as in the previous section and derive equations for the long-wavelength overdensity and momentum
\begin{align}
\partial_\tau \delta_l + \frac{1}{a} \partial_i \pi^i_l &= 0,\label{eq:eom1}\\
\partial_\tau \pi^i_l +\frac{1}{a} \partial_j \left(\frac{\pi^i_l \pi^j_l}{\rho_l}\right)+ a \rho_l \partial^i \phi_l &= -\partial_j \uptau^{ij}.\label{eq:eom2}
\end{align}
The effect of the short-wavelength modes is encoded in the stress tensor $ \uptau^{ij}$, which can be parametrized by all possible interactions of the smoothed fields consistent with the symmetries of the system. In \Sec{sec:stress} we describe in detail the construction and the properties of the stress tensor up to NNLO.

As in SPT, it is convenient to rewrite the equations of motion in terms of the velocity. As noted already in Refs.~\cite{2014JCAP...03..006M,2015arXiv150907886A}, the relation $\pi^i = a\rho v^i$ is not preserved under smoothing, as
\beq 
\int d^3 x'  \,W_\Lambda (\vec{x} - \vec{x}')\, \rho(\vec{x}') v(\vec{x}') \neq  \int d^3 x'   \,W_\Lambda (\vec{x} - \vec{x}')\, \rho(\vec{x}')   \int d^3 x''   \,W_\Lambda (\vec{x} - \vec{x}'')\, v(\vec{x}''). 
\eeq 
Thus, we define \beq 
\pi^i_l = a (\rho_l v^i_l  + \bar{\rho} \Sigma^i),
\eeq
where the heat conduction $\Sigma^i$ parametrizes additional terms arising from the smoothing of the composite operator. With this substitution, Eqs.~\eqref{eq:eom1} and \eqref{eq:eom2} become 
\begin{align*}
\partial_\tau \delta_l + \theta_l &= S_\alpha - \partial_i \Sigma^i,\numberthis\label{deltaeqn1}\\
 \partial_\tau \theta_l +  \mathcal{H} \theta_l + \frac{3}{2} \mathcal{H}^2 \delta_l &= S_\beta -\partial_i \left(\frac{\partial_j \uptau^{ij}}{1+\delta_l}\right)\\
 &\quad - \partial_i \left(\frac{\partial_\tau \Sigma^i + \mathcal{H} \Sigma^i - v^i_l \partial_j \Sigma^j +\partial_j(v_l^i\Sigma^j +v_l^j \Sigma^i)}{1+\delta_l}\right),\numberthis\label{thetaeqn1} \\
\partial_\tau \omega^i_l +  \mathcal{H} \omega^i_l &= S_\gamma -\epsilon^{ijk} \partial_j \left(\frac{\partial_m \uptau^{km}}{1+\delta_l}\right)\\
&\quad -\epsilon^{ijk} \partial_j \left(\frac{\partial_\tau \Sigma^k + \mathcal{H} \Sigma^k - v_l^k \partial_m \Sigma^m +\partial_m(v_l^k\Sigma^m +v_l^m \Sigma^k)}{1+\delta_l}\right).\numberthis\label{omegaeq1}
\end{align*} 
Here, $S_\alpha,~S_\beta,$ and $S_\gamma$ indicate collectively all the SPT mode-coupling functions appearing on the right-hand side of Eqs.~(\ref{deltaeqn}-\ref{omegaeq}), and we have reabsorbed a factor of $a\bar{\rho}$ into the definition of $\uptau^{ij}$. 

\subsection{Perturbative Solution}

The equations laid out in Eqs.~(\ref{deltaeqn1}-\ref{omegaeq1}) constitute the complete set of equations in the EFT of LSS. Once a parametrization of $\uptau^{ij}$ and $\Sigma^i$ in terms of the long-wavelength fields is provided, they can be solved perturbatively by expanding the fields in powers of the linear density perturbation.
We impose the standard perturbative ansatz for the growing modes:
\begin{align}
\delta_l(\vec{k},\tau)&=\sum_{n=1}^\infty\left( D^n(\tau)\,\delta_n(\vec{k})+\varepsilon\,D^{n+2}(\tau)\,\tilde{\delta}_n(\vec{k})\right),\label{eq:ansatzd}\\
\theta_l(\vec{k},\tau)&=-\mathcal{H}f(\tau)\sum_{n=1}^\infty\left( D^n(\tau)\,\theta_n(\vec{k})+\varepsilon\,D^{n+2}(\tau)\,\tilde{\theta}_n(\vec{k})\right),\label{eq:ansatzt}\\
\omega_l^i(\vec{k},\tau)&=-\mathcal{H}f(\tau)\sum_{n=2}^\infty \varepsilon\,D^{n+2}(\tau)\,\tilde{\omega}_{n}^i(\vec{k}),\label{eq:ansatzw}
\end{align}
where we have assumed that the SPT part of vorticity can be neglected, and where each of the fields on the right-hand side can be written in terms of $n$ powers of the linear density perturbation which is small on large scales, $\delta_1\ll 1$. The first term on the right hand side of the first two equations contains the standard SPT perturbative ansatz, while an $\varepsilon$ is introduced to track the leading EFT corrections, which are of order $\mathcal{O}(k^2/k^2_\text{NL})$. The EFT source terms $\uptau^{ij}$ and $\Sigma^i$ can also be expanded both in powers of $\delta_1$ and in powers of $\varepsilon$, starting at $\mathcal{O}(\varepsilon)$.
In the above ansatz, $D(\tau)$ is the linear growth function, $f(\tau)=\,d\ln D(\tau)/\mathcal{H} d\tau$, and we assume $f(\tau) = \sqrt{\Omega_m}$. For an EdS universe ($\Omega_m=1$) the equations of motion are fully separable and the solution can always be written in the form of Eqs.~(\ref{eq:ansatzd}-\ref{eq:ansatzw}), with $D(\tau)=a(\tau)$ and $f(\tau)=1$. Even though for a $\Lambda{\rm CDM}$ universe the time-dependence should be recomputed at each order in perturbations, it has been shown that Eqs.~(\ref{eq:ansatzd}-\ref{eq:ansatzw}) are a good approximation. For the one-loop power spectrum and bispectrum, for example, the approximation is valid up to corrections of $\order(1\%)$~\cite{2008PThPh.120..549T,2012JHEP...09..082C,2015JCAP...05..007B}. The exponent $n+2$ for the EFT time-dependence is chosen such that the EFT contributions have the same time-dependence as the loop contributions from SPT.
 
With the ansatz in Eqs.~(\ref{eq:ansatzd}-\ref{eq:ansatzw}) one can solve Eqs.~(\ref{deltaeqn1}-\ref{omegaeq1}) order by order. At each perturbative order $n$, the $\mathcal{O}(\varepsilon^0)$ equations will produce the SPT solution and the $\mathcal{O}(\varepsilon)$ will determine the leading EFT correction. Each field in Eqs.~(\ref{eq:ansatzd}-\ref{eq:ansatzw}) can be written as a convolution of $n$ linear density perturbations with kernels as
 \begin{align*}
\begin{pmatrix} 
\delta_n(\vec{k}) \\ 
\theta_n(\vec{k}) \\
\tilde{\delta}_n(\vec{k}) \\
\tilde{\theta}_n(\vec{k})\\
\tilde{\omega}_n^i(\vec{k})  
\end{pmatrix}& = 
\int \dbar^{\,3}q_1 ... \ \dbar^{\,3}q_n \, \begin{pmatrix} F_n(\vec{q}_1,...,\vec{q}_n) \\ G_n(\vec{q}_1,...,\vec{q}_n) \\ \widetilde{F}_n(\vec{q}_1,...,\vec{q}_n) \\
\widetilde{G}_n(\vec{q}_1,...,\vec{q}_n)\\
\widetilde{G}_{n}^{\omega i}(\vec{q}_1,...,\vec{q}_n)
\end{pmatrix}
(2 \pi)^3 \delta_D\left(\vec{k}-\sum_{i=1}^n{\vec{q}_i}\right) \delta_1(\vec{q}_1)... \delta_1(\vec{q}_n),
 \numberthis \label{eq:kernels}\end{align*} 
where $\dbar^{\,3}q\equiv d^{\,3}q/(2\pi)^3$. The SPT kernels $F_n$ and $G_n$ can be determined from well-known recursion relations \cite{1986ApJ...311....6G, 1994ApJ...431..495J, 2002PhR...367....1B}, and we have described the general form of the EFT kernels up to $n=3$ in our previous paper \cite{2015arXiv151207630B}.  In Sec.~\ref{sec:stress}, we will present a more detailed discussion on the construction of the EFT sources up to NNLO and the derivation of the EFT kernels, which are collected in Appendix~\ref{app:kernels}.

\subsection{Heat Conduction Terms and Vorticity}
\label{sec:heat}
Vorticity is usually neglected in SPT, which means that only the terms in Eqs.~(\ref{deltaeqn}) and~(\ref{thetaeqn}) involving the $\alpha$ and $\beta$ kernels are considered. This is justified from Eq.~\eqref{omegaeq}: at any given perturbative order, the vorticity will decay relative to the velocity divergence. For instance, the source term for the linear perturbations to $\omega$ enters only at second order, meaning that the leading order behavior is for the linear vorticity to be damped away by the Hubble drag term. Beyond the linear regime, one could obtain growing vorticity modes, but the sources will always contain powers of the linear-order vorticity and hence be suppressed by factors of $1/a(\tau)$ relative to the terms sourcing growing modes of $\delta$ and $\theta$. Even primordial vorticity is damped away by the expansion of the universe within the SPT framework. 

However, we know that on some scale there is vorticity which spins up dark matter halos \cite{2015MNRAS.446.2744L}, and there could be feedback between this vorticity and large scale density modes. As seen from Eq.~\eqref{omegaeq1}, the stress tensor and heat conduction terms of the EFT of LSS source a non-decaying vorticity at NLO. In this section, we illustrate how this induced vorticity is relevant for computing the EFT contributions to the trispectrum.

Let us begin with a discussion of a field redefinition for the velocity that is commonly adopted in the literature. If we are interested in calculating only correlators of the density perturbation $\delta_l$, we can reabsorb the heat conduction terms $\Sigma^i$ into a redefinition of the velocity \cite{2014JCAP...03..006M,2015arXiv150907886A}, 
\beq
\label{eq:vpi}
v_\pi^i=v^i_l+\frac{\Sigma^i}{1+\delta_l}.
\eeq
If one uses $v_\pi$, the SPT relation between velocity and momentum is preserved, $\pi_l^i=a\rho_lv_\pi^i$, and the new set of equations is simpler as it only involves the stress tensor $\uptau^{ij}$ as an effective source,
\begin{align}
\partial_\tau \delta_l + \theta_\pi &= S_\alpha,\numberthis\label{deltaeqn2}\\
 \partial_\tau \theta_\pi +  \mathcal{H} \theta_\pi + \frac{3}{2} \mathcal{H}^2 \delta_l &= S_\beta -\partial_i \left(\frac{\partial_j \uptau^{ij}}{1+\delta_l}\right), \numberthis\label{thetaeqn2} \\
\partial_\tau \omega^i_\pi +  \mathcal{H} \omega^i_\pi &= S_\gamma -\epsilon^{ijk} \partial_j \left(\frac{\partial_b \uptau^{kb}}{1+\delta_l}\right),\numberthis\label{omegaeq2}
\end{align}
where $\theta_\pi$ and $\omega_\pi^i$ indicate the divergence and vorticity of $v_\pi$, respectively. Equations (\ref{deltaeqn2}-\ref{omegaeq2}) constitute the set of equations which is usually used in the EFT of LSS literature. Here, we would like to point out that, consistent with the field redefinition, even if one were to instead use Eqs.~(\ref{deltaeqn1}-\ref{omegaeq1}), all the terms involving $\Sigma^i$ cancel when calculating $\delta_l$ correlators. In particular, 
\beq
\widetilde{F}_n(\vec{k}_1,...,\vec{k}_n)|_{\{\delta_l,v_l\}}=\widetilde{F}_n(\vec{k}_1,...,\vec{k}_n)|_{\{\delta_l,v_\pi\}},
\eeq
where the $\widetilde{F}_n$ kernels have been introduced in Eq.~\eqref{eq:kernels}, and the labels $\{\delta_l,v_l\}$ and $\{\delta_l,v_\pi\}$ denote the bases of Eqs.~(\ref{deltaeqn1}-\ref{omegaeq1}) and (\ref{deltaeqn2}-\ref{omegaeq2}), respectively. 

Before illustrating this point with a specific example, let us comment on the role of vorticity in this field redefinition. As mentioned above, in the EFT of LSS, both the stress tensor and heat conduction terms  serve as sources for vorticity. In particular, the first non-vanishing contribution arises at NLO (the leading order does not contribute as it is curl-free) and feeds back into the continuity and Euler equations starting at NNLO, which is the relevant order for the trispectrum at one loop. For the purpose of reabsorbing the cutoff dependence of SPT diagrams for density correlators, one can still neglect the vorticity of the smoothed velocity and set $\omega_l^i =0$ by imposing a cancellation of the two sources in Eq.~\eqref{omegaeq1}. The $\widetilde{F}_3$ kernel will still remain independent of this choice (while the $\widetilde{G}_3$ kernel will not). However, it should be noticed that if one uses the basis of Eqs.~(\ref{deltaeqn2}-\ref{omegaeq2}), even if the vorticity of the smoothed velocity $\omega_l^i$ is assumed to be zero, then $\omega_\pi^i\neq 0$. This can easily be seen from Eq.~\eqref{omegaeq2}, where vorticity is now sourced by the stress tensor which contains, in general, non-curl-free operators, or equivalently from Eq.~\eqref{eq:vpi} where $\omega_\pi^i$ would receive a contribution from the curl of $\Sigma^i/(1+\delta_l)$.

We will now illustrate the cancellation of the $\Sigma^i$ terms and the role of vorticity with a simplified example. Let us assume the following form for the stress tensor and heat conduction terms
\begin{align*}
\uptau^{ij}&={\cal H}^2 f(\tau)^2 D(\tau)^2\left[ c_1\delta^{ij}\delta_l(\vec{k})+c_2\int d^3 q \, \frac{q^iq^j}{q^2}\,\delta_l(\vec{q})\delta_l(\vec{k}-\vec{q})\right],\numberthis\\
\Sigma^i&={\cal H} f(\tau)^2 D(\tau)^2\left[\chi_1k^i\delta_l(\vec{k})+\chi_2\int d^3q\, \frac{\left((\vec{k}-\vec{q})\cdot\vec{q}\right)q^i}{q^2}\,\delta_l(\vec{q})\delta_l(\vec{k}-\vec{q})\right].\numberthis
\end{align*}
One can show from Eqs.~(\ref{deltaeqn1}-\ref{omegaeq1}), without making any field redefinitions, that
\begin{align*}
\widetilde{F}_1(\vec{k})& = \widetilde{F}_1(\vec{k})|_{\chi_{1,2}=0},\numberthis\\
\widetilde{F}_2(\vec{k}_1,\vec{k}_2)& = \widetilde{F}_2(\vec{k}_1,\vec{k}_2)|_{\chi_{1,2}=0},\numberthis\\
\widetilde{F}_3(\vec{k}_1,\vec{k}_2,\vec{k}_3)& = \widetilde{F}_3(\vec{k}_1,\vec{k}_2,\vec{k}_3)|_{\chi_{1,2}=0},\numberthis
\end{align*}
because terms involving $\chi_{1,2}$ cancel.
Thus, correlators of $\delta_l$ are indeed independent of the heat conduction terms, \emph{i.e.} they do not depend on the definition of the velocity used. We again emphasize that $\widetilde{F}_3(\vec{k}_1,\vec{k}_2,\vec{k}_3)$ above contains contributions from vorticity. Equivalently, one can consistently neglect vorticity in the $\{\delta_l,v_l\}$ basis. However, this implies that the EFT sources on the right-hand side of Eq.~\eqref{omegaeq1} must cancel, requiring $\chi_2=-2/9 c_2$. Still, the $\widetilde{F}_3$ kernel is independent of this choice,
\beq
\widetilde{F}_3(\vec{k}_1,\vec{k}_2,\vec{k}_3)=\widetilde{F}_3(\vec{k}_1,\vec{k}_2,\vec{k}_3)|_{\omega_l^i=0,\chi_2=-\frac{2}{9} c_2}.
\eeq
To summarize, one can choose to work with $v_\pi^i$ and neglect heat conduction terms, in which case vorticity must be included; equivalently, one can work with $v_l^i$  and neglect vorticity but in this case heat conduction terms must be included.

\section{EFT of LSS Stress Tensor}
\label{sec:stress}

In our analysis, we choose to work with $v_\pi^i$ such that the relevant equations of motion are given by Eqs.~(\ref{deltaeqn2}-\ref{omegaeq2}). Working in terms of this velocity field (related to the physical velocity field $v_l^i$ through the field redefinition in Eq.~\eqref{eq:vpi}) is valid since we are interested in computing the trispectrum of the density field only, and convenient since heat conduction terms are not required in this basis. To make the notation less cumbersome, we drop the subscripts on the fields and note that for the rest of the paper we are always working with $\delta_l$ and $v_\pi^i$. 
In Sec.~\ref{sec:symmetries}, we briefly review the relevant symmetries imposed on the stress tensor and the basic building blocks from which it is constructed. In Sec.~\ref{sec:TNL}, we discuss the time non-locality of the stress tensor, and provide an equivalent formulation in terms of operators that are local in time but not space (see Eq.~\eqref{eq:TNLresult}). These are then expressed in terms of convective derivatives (see Eq.~\eqref{eq:TNLresult2}), which in Sec.~\ref{sec:local} we then show are redundant through NNLO. Thus, within the Eulerian framework, it is sufficient to consider trispectrum counterterms that are local in time and space with no convective derivatives, and an explicit prescription for the counterterms is given in Sec.~\ref{sec:tensor}. Furthermore, our methods suggest that this is true for density correlators at all orders in Eulerian perturbation theory. In Sec.~\ref{sec:kernels}, we then derive the form of the EFT kernels up to NNLO.

\subsection{Symmetries and Building Blocks}\label{sec:symmetries}

The cosmological principle states that there is not a special point or direction in our universe, and hence the laws of physics must be the same everywhere at a given conformal time. Since the expansion of the universe breaks time translation invariance, any coefficients that appear in the equations of motion are at most functions of time. Additionally, this means that all physical quantities, such as the density and velocity, must be statistically homogeneous and isotropic at a given conformal time.

Since we are working well below the Hubble scale where relativistic corrections become important, the relevant symmetry is Galilean invariance. The equations of motion must be invariant under the transformation $x \rightarrow x' = x+ n(\tau)$, where $n$ can only depend on time. Under this Galilean boost, the physical quantities in our problem transform as 
\begin{align}
& \rho(x)\rightarrow \rho(x')\,, \nonumber \\
& v^i(x)\rightarrow v^i(x') - \partial_\tau n^i\,, \nonumber\\
& \partial_\tau \rightarrow \partial_{\tau} +  \partial_\tau \vec{n}\cdot \vec{\partial}\,, \nonumber \\
&\phi(x) \rightarrow \phi(x') + \mathcal{H} \vec{x}\cdot \left( \partial_\tau \vec{n}+ \partial_\tau^2 \vec{n} \right)  \,,
\end{align}
where the first three transformations are the usual Galilean transformations in comoving coordinates, and the transformation of the gravitational potential $\phi(x)$ is deduced from the invariance of the Poisson equation under boosts in an expanding space. 
Note that the convective derivative $D_\tau = \partial_\tau + \vec{v} \cdot \vec{\partial}$ is invariant under these transformations.
It is straightforward to check that the equations for mass and momentum conservation  (Eqs.~(\ref{eq:eom1}-\ref{eq:eom2})) are also invariant as long as the stress tensor $\uptau_{ij}$ and heat conduction $\Sigma_i$ are invariant by construction. To this end, a set of Galilean invariant building blocks are given by
\beq\label{eq:bblocks}
\partial_i \partial_j \phi\, , \quad \partial_i \partial_j \phi_v \equiv \partial_i v_j \,,
\eeq
as well as their spatial ($\partial_k$) and convective ($D_\tau$) derivatives. Below, we denote the potentials $\phi$ and $\phi_v$ generically as $\Phi$. For the present case, we neglect vorticity when constructing the stress tensor since we work at $\mathcal{O}(\varepsilon^1)$, while including vorticity in the stress tensor would yield $\mathcal{O}(\varepsilon^2)$ contributions. Note that $\partial_i D_\tau \partial_j \Phi$ is also an invariant but is equal to 
$D_\tau \partial_i \partial_j \Phi + \left(\partial_i v_k\right) \partial_k \partial_j \Phi $, 
and is therefore redundant. 

\subsection{Non-locality in Time}\label{sec:TNL}
The hierarchy of scales that allows us to define the EFT of LSS, {\it e.g.}, through smoothing as in Eq.~\eqref{eq:EFTwindow}, only applies for length scales and not time scales. 
In particular, the linear equations of motion are scale invariant, which means that different modes grow at the same slow rate, giving rise to the notion of a universal linear growth function, $D(\tau)$.  Since even small scales evolve slowly, they cannot be integrated out all at once (unlike in quantum field theory); their coupling to large scale modes evolves with time.  In other words, there is memory in the system, and one has to consider the entire history of a given mode to see the cumulative effects of mode coupling at a given conformal time \cite{2014JCAP...07..057C}.

Tracking the evolution of large-scale modes over time can be done through coordinates $\vec{x}_{fl}$ that are comoving with the displaced fluid elements inside a given mode. We thus write the fluid element's position at some earlier time $\tau'$ as a recursive time-ordered expansion about its current position at time $\tau$, 
 \beq 
 \vec{x}_{fl}(\tau, \tau') = \vec{x} - \int_{\tau'}^\tau d \tau'' \vec{v}(\vec{x}_{fl}(\tau, \tau''), \tau''). 
 \eeq
We then express the stress tensor in terms of the history of this fluid element's position over time, that is
 \beq   \label{eq:TNLstress}
{\uptau}_{ij}=    \int d \tau ' K (\tau, \tau') {\uptau}^{\rm loc}_{ij}(\vec{x}_{{fl}}, \tau') \,, 
 \eeq 
where ${\uptau}_{ij} $ is the stress tensor appearing in Eqs.~(\ref{deltaeqn2}-\ref{omegaeq2}), and contains the memory effects through the kernel $K$, while $\uptau^{\rm loc}_{ij}(\vec{x}_{{fl}}, \tau')$ is a stress tensor that is local in time and space, constructed from the building blocks in Eq.~\eqref{eq:bblocks}, and evaluated along the fluid trajectory $\vec{x}_{fl}$. While this form is daunting because it is non-local in time, we can perturbatively write quantities at the position $\vec{x}_{fl}$ in terms of an expansion about fixed coordinates ({\em e.g.} the fluid's current position $\vec{x}$). Working in Eulerian space, we can expand as follows
 \begin{align*} \label{eq:TNLexpand}
\uptau^{\rm loc}_{ij}(\vec{x}_{{fl}}, \tau')= & \uptau^{\rm loc}_{ij}(\vec{x}, \tau' ) - \partial_k \uptau^{\rm loc}_{ij}(\vec{x}, \tau' ) \int_{\tau'}^{\tau} d\tau'' v^k(\vec{x}, \tau'' ) \\
&+ \partial_k \uptau^{\rm loc}_{ij}(\vec{x}, \tau' ) \int_{\tau'}^{\tau} d \tau'' \partial_b v^k(\vec{x}, \tau'') \int_{\tau''}^{\tau} d \tau''' v^b(\vec{x}, \tau''' )\\
& +\frac{1}{2} \partial_k \partial_b \uptau^{\rm loc}_{ij}(\vec{x}, \tau' ) \int_{\tau'}^{\tau} d \tau'' v^k(\vec{x}, \tau'' ) \int_{\tau'}^{\tau} d \tau''' v^b(\vec{x}, \tau''' )+\ldots \,, \numberthis
\end{align*}
where the ellipsis denotes terms with more velocity fields.
Upon plugging Eq.~\eqref{eq:TNLexpand} into \eqref{eq:TNLstress}, expanding the fields in terms of the perturbative ansatz, and identifying the coefficients
\begin{align}
c_r(a) &= \int { da'' \over a'' {\cal H}(a'') } K(a,a'') \left( a'' / a  \right)^r, \nonumber \\
c_{rs}(a) &=  {1 \over s} \left(  c_r(a) - c_{r+s}(a) \right),  \nonumber \\
c_{rst}(a) &=  {1 \over st}  \left( c_r(a) - c_{r+s}(a) - c_{r+t}(a) + c_{r+s+t}(a) \right)  \nonumber \\
&= \left( {1 \over s(s+t)} c_r(a) - {1 \over st} c_{r+s}(a) + {1 \over t(s+t)} c_{r+s+t}(a)  \right) + ( t \leftrightarrow s )\,,
\label{eq:TNLcoeffs}
\end{align}
we find
\begin{align*} \label{eq:TNLresult}
 {\uptau}_{ij} &=   \sum_{r=1}^\infty c_r(a) \uptau_{ij,\,r}^{{\rm loc} }(\vec{x},a)  + \sum_{r,s=1}^\infty c_{rs}(a)  v^k_{s}(\vec{x},a)  \partial_k \uptau_{ij,\,r}^{{\rm loc} }(\vec{x},a)  \\
& + \frac12 \sum_{r,s,t=1}^\infty c_{rst}(a)    v^b_{t}(\vec{x},a)    \partial_b  \left( v^k_{s}(\vec{x},a) \partial_k \uptau_{ij,\,r}^{{\rm loc} } (\vec{x},a) \right) + \dots \, . \numberthis
\end{align*}
Here, the subscripts on the stress tensor and velocity track their perturbative order. The first and second terms in Eq.~\eqref{eq:TNLexpand} lead to the first and second terms in Eq.~\eqref{eq:TNLresult}, while the last two terms in Eq.~\eqref{eq:TNLexpand} combine (using the identity for $c_{rst}(a)$ in Eq.~\eqref{eq:TNLcoeffs}) to the last term in Eq.~\eqref{eq:TNLresult}. In the local-in-time case where $K(a,a') \propto \delta(a-a')$, the coefficient $c_r(a)$ is independent of $r$, while the rest of the coefficients vanish. 

The result Eq.~\eqref{eq:TNLresult} shows that the time non-local stress tensor can be written as a set of local in time operators involving the velocity field, that are weighted (by the coefficients in Eq.~\eqref{eq:TNLcoeffs}) according to their time evolution ({\em i.e.}, $a$-scalings). 
We can further express Eq.~\eqref{eq:TNLresult} in terms of convective derivatives through the construction
\begin{align}\label{eq:TNLresult2}
{\uptau}_{ij} = \sum_{n=1}^\infty d_n(a) \left( D_\tau \over {\cal H} a \right)^n \uptau_{ij}^{\rm loc}(x,a)   \,, \quad  d_n(a) = \sum_{k=1}^n {(-1)^{n+k} a^n c_k(a) \over k! (n-k)! }\,, 
\end{align}
where the normalization ${1/{\cal H} a}$ is for convenience and dimensional consistency.  We find that the terms with zero, one, and two velocity fields are 
\begin{align}
&{\uptau}_{ij} = \sum_{n=1}^\infty d_n(a)  \left( \partial_\tau \over {\cal H} a \right)^n \uptau_{ij}^{\rm loc}  
+ \sum_{n=1}^\infty  \sum_{m=0}^{n-1} d_n(a)  \left( \partial_\tau \over {\cal H} a \right)^{n-m-1}\left( v^k  \partial_k  \left( \partial_\tau \over {\cal H} a \right)^m \uptau_{ij}^{\rm loc}\right)  \nonumber \\
&+ \sum_{n=1}^\infty \sum_{r=0}^{n-m-2}   \sum_{m=0}^{n-2}    d_n(a)   \left( \partial_\tau \over {\cal H} a \right)^{n-m-r-2}  \left(v^b  \partial_b \left( \partial_\tau \over {\cal H} a \right)^r \left(v^k \partial_k\left( \partial_\tau \over {\cal H} a \right)^m  \uptau_{ij}^{\rm loc} \right)\right)
 + ... \,,
\end{align}
which, after some algebra and with the above choice for the coefficients $d_n(a)$, reduces exactly to Eq.~\eqref{eq:TNLresult}. 
While the result in Eq.~\eqref{eq:TNLresult} involves structures that are not spatially local ($v^k = {\partial^k \over \partial^2} \theta$), the result in Eq.~\eqref{eq:TNLresult2}, involving a series of convective derivatives, is local both in time and space.
This establishes a simple formulation of time non-locality in the Eulerian framework through NNLO. The construction may be extended to higher orders by including the terms with more velocity fields.
Such a connection between the Eulerian formulation of time non-locality and convective derivatives is of course immediate from a Taylor expansion of $\uptau_{ij}^{\rm loc}(\vec{x}_{fl}, \tau^\prime)$ about the final time, and is valid when the characteristic time scale of the memory kernel $K$ is much smaller than a Hubble time \cite{2015JCAP...05..007B}.

\subsection{Convective Derivatives}\label{sec:local}

Using the stress tensor ${\uptau}_{ij}$ in Eq.~\eqref{eq:TNLresult2} that is local in time and space, let us now construct the term appearing on the right-hand side of Eqs.~(\ref{deltaeqn2}-\ref{omegaeq2}), 
\beq\label{eq:st1}
\partial_i \left( \, {1 \over 1+ \delta} \, \partial_j { \uptau}_{ij}\right) = \partial_i  \partial_j { \uptau}_{ij} - \partial_i \left(\delta \partial_j { \uptau}_{ij} \right)+ \partial_i \left( \delta^2 \partial_j { \uptau}_{ij} \right)+ \dots \,,
\eeq 
where we have expanded $1/(1+\delta)$, keeping only terms that contribute through NNLO. The enumeration of terms may be further organized into the number of fields (not counting convective derivatives), as well as the number of convective derivatives appearing in the operator. 
This is summarized in Table~\ref{tab:terms}. Note that, for enumerating operators through a given order, we may count the convective derivative as having one field, \emph{i.e.} $D_\tau \sim v \cdot \partial$, since $\partial_\tau$ alone does not yield independent operators.

\begin{table}[t]
\centering
\begin{tabular}{|c|c|c|c|} 
\hline
\# fields &$\partial_i  \partial_j {\uptau}_{ij}$&$\partial_i( \delta \partial_j {\uptau}_{ij} )$&$\partial_i ( \delta^2 \partial_j {\uptau}_{ij})$\\
\hline
$1$ \quad&0, 1, 2 & 0, 1 & 0\\
\hline
$2$ \quad& 0, 1 & 0&  \\
\hline
3 \quad & 0 &  & \\
\hline
\end{tabular}
\caption{The structure of terms in the stress tensor ${\uptau}_{ij}$ relevant through NNLO, \emph{i.e.} $\order(\delta_1^3)$. The left-most column gives the number of fields in the operator, not counting convective derivatives.  The remaining columns correspond to each term in Eq.~\eqref{eq:st1}, and the entries in each cell denote the possible number of convective derivatives appearing in the operator. The empty cells are not relevant through NNLO.\label{tab:terms}}
\end{table}

We begin by considering the terms with convective derivatives, given by
\begin{align}\label{eq:timelist}
\partial_i \partial_j  \left(D_\tau \partial_k \partial_l \Phi\right), \quad \partial_i \partial_j \left( D_\tau^2 \partial_k \partial_l \Phi \right), \quad \partial_i  \left( \delta \partial_j D_\tau \partial_k \partial_l \Phi\right),  \quad \partial_i \partial_j \left(\left( \partial_k \partial_l \Phi^\prime \right) D_\tau \partial_m \partial_n \Phi\right) ,
\end{align}
where all unique contractions of the indices $\{i,j,\dots\}$ are considered, and $\Phi, \Phi^\prime$ denote distinct gravitational or velocity potentials. Note that, since $D_\tau$ counts practically as having one field, the last three operators in Eq.~\eqref{eq:timelist} are already at $\order(\delta_1^3)$, and thus we may take $\Phi=\Phi^\prime= \phi_v$. 
This choice of $\phi_v$ over $\phi$ is convenient since recursively applying the bare equations of motion, 
\beq
D_\tau^{m+1} \partial_j v_i  + D_\tau^{m} \left(\left( \partial_j v_k \right)\left( \partial_k v_i \right)\right)+ D_\tau^{m}\left( \mathcal{H} \partial_j v_i \right)+ D_\tau^{m} \partial_j \partial_i \phi = 0\,,
\eeq
allows us to replace $D_\tau^n \partial_j v_i$ with terms having no convective derivatives, except on $\partial_i \partial_j \phi$. In particular, the replacement of $D_\tau \partial_j v_i$ involves no convective derivatives, while the replacement of $D_\tau^2 \partial_j v_i$ involves $D_\tau \partial_i \partial_j \phi$. This shows that all operators in Eq.~\eqref{eq:timelist} with $\Phi, \Phi^\prime = \phi_v$ are redundant, leaving us with $\partial_i \partial_j\left( D_\tau \partial_k \partial_l \phi\right)$, which has two possible contractions:
\beq\label{eq:timelist2}
\partial^2 \left(D_\tau \partial^2 \phi\right) , \quad \partial_i \partial_j \left(D_\tau \partial_i \partial_j \phi \right) .
\eeq 
We find that these are also redundant with operators having no convective derivatives by using the equation of motion $D_\tau \delta = - \delta \theta$, and the identity 
\beq
\partial_j \left(D_\tau \partial_i \partial_j \Phi\right) = \partial_i \left(D_\tau \partial^2 \Phi \right)+ \partial_i\left( \left( \partial_j v_k\right) \left(\partial_k \partial_j \Phi \right) \right)-  \partial_j \left(\left( \partial_i v_k \right) \left(\partial_k \partial_j \Phi \right)\right) .
\eeq
We can therefore neglect operators with convective derivatives through NNLO.  

\subsection{Local Stress Tensor}\label{sec:tensor}
To summarize the previous two subsections, we have shown explicitly that non-locality in time is equivalent to a series of operators that are local in time and space, and involve convective derivatives (see Eq.~\eqref{eq:TNLresult2}), and furthermore that the operators with convective derivatives are redundant by the equations of motion. The operators with no convective derivatives have the following structure:
\begin{align}
\partial_i \, \left({1 \over 1+ \delta} \, \partial_j {\uptau}_{ij}\right) \supset & \ \partial^4 \Phi \,, \quad \partial_i \left(\delta \partial_i \partial^2 \Phi \right), \quad  \partial_i \left(\delta^2 \partial_i \delta\right), \nonumber \\
& \ \partial_i \partial_j \left(\left( \partial_k \partial_l \Phi \right)  \left(\partial_m \partial_n \Phi^\prime \right)\right), \quad \partial_i\left( \delta \partial_j \left(\left( \partial_k \partial_l \phi \right) \left( \partial_m \partial_n \phi \right)\right)\right), \nonumber \\
& \ \partial_i \partial_j \left(\left( \partial_k \partial_l \phi \right) \left(\partial_m \partial_n \phi\right) \left( \partial_o \partial_p \phi \right)\right),
\end{align}
where all unique contractions of the indices $\{i,j,\dots\}$ are considered, and we have used the gravitational potential for operators with three fields.
Let us specify these operators explicitly by 
constructing $\partial_i \uptau_{ij}$. Writing all possible terms, we have in Fourier space:
\begin{align*}\label{eq:stresstensor}
&k_i { \uptau}^{ij}  = \, \bar{c}_s^\delta k^j  \delta({\vec k}) + {\bar{c}_s^\theta\over \mathcal{H}f}   k^j \theta({\vec k})\\
&\quad + \int d\vec{q} \sum_{n=1}^4 \Big[\bar{c}_n^{\delta\delta}  \delta({{\vec q}})\delta(\vec{k} - {{\vec q}}) +{\bar{c}_n^{\theta\theta}\over \mathcal{H}^2f^2}  \theta({{\vec q}})\theta(\vec{k} - {{\vec q}})\\
&\qquad \qquad + {\bar{c}_n^{\delta\theta}\over \mathcal{H}f}   \delta({{\vec q}}) \theta(\vec{k} - {{\vec q}}) 
+ {\bar{c}_n^{\theta \delta}\over \mathcal{H}f}  \theta({{\vec q}}) \delta(\vec{k} - {{\vec q}} )
 \Big]  k_i e^{ij}_n (\vec{q}, \vec{k} - \vec{q}) 
\\
&\quad  +  \int d\vec{q}_1 d\vec{q}_2  \sum_{n=1}^{10} \bar{c}_n^{\delta\delta\delta}   \delta({{\vec q}_1}) \delta({{\vec q}_2}) \delta(\vec{k} - {{\vec q}_1} - \vec{q}_2)  k_i E^{ij}_n (\vec{q}_1, \vec{q}_2, \vec{k} - {{\vec q}_1} - \vec{q}_2) \,. \numberthis 
\end{align*} 
Here the functions $e_n^{ij}$ and $E_n^{ij}$ account for all possible contractions of the indices, and are given by
\begin{align*}\label{eq:operators}
 E_1^{ij} (\vec{q}_1, \vec{q}_2, \vec{q}_3)&= e_1^{ij} (\vec{q}_1, \vec{q}_2) =\delta^{ij} \,, &
E_2^{ij} (\vec{q}_1, \vec{q}_2, \vec{q}_3)&= e_2^{ij} (\vec{q}_1, \vec{q}_2) = {q_1^i q_1^j \over q_1^2} \,, \\
E_3^{ij} (\vec{q}_1, \vec{q}_2, \vec{q}_3)&= e_3^{ij} (\vec{q}_1, \vec{q}_2) = {q_1^{\{ i} q_2^{j\}} q_1^a q_2^a \over q_1^2 q_2^2} \,, &
E_4^{ij} (\vec{q}_1, \vec{q}_2, \vec{q}_3) &= e_4^{ij} (\vec{q}_1, \vec{q}_2) = {\delta^{ij}  (q_1^a q_2^a)^2 \over q_1^2 q_2^2} \,, \\
E_5^{ij} (\vec{q}_1, \vec{q}_2, \vec{q}_3)&= {q_1^i q_1^j (q_2^a q_3^a)^2 \over q_1^2 q_2^2 q_3^2 }  \,, &
E_6^{ij} (\vec{q}_1, \vec{q}_2, \vec{q}_3)&= {q_1^{\{ i} q_2^{j\}} q_1^a q_3^a q_2^b q_3^b \over q_1^2 q_2^2 q_3^2 }\,, \\
E_7^{ij} (\vec{q}_1, \vec{q}_2, \vec{q}_3)&= {\delta^{ij} q_1^a q_2^a q_2^b q_3^b q_3^c q_1^c \over q_1^2 q_2^2 q_3^2 } \, , &
E_8^{ij} (\vec{q}_1, \vec{q}_2, \vec{q}_3)&=   { \epsilon^{\{ iab} \epsilon^{jcd \}} q_1^a q_1^c q_2^b q_2^d \over q_1^2 q_2^2  }  \,, \\ 
E_9^{ij} (\vec{q}_1, \vec{q}_2, \vec{q}_3)&=   {\epsilon^{\{ iab} \epsilon^{jcd \}} q_1^a q_1^e q_2^c q_2^e q_3^b q_3^d \over q_1^2 q_2^2 q_3^2 }\,, &
E_{10}^{ij} (\vec{q}_1, \vec{q}_2, \vec{q}_3)&=  {\delta^{ij} (\epsilon^{abc}  q_1^a q_2^b  q_3^c )^2 \over q_1^2 q_2^2 q_3^2 } \,,   \numberthis
\end{align*}
where $\{ \ \}$ denotes symmetrization in the indices $i,j$. For each operator above, we have introduced a coefficient $\bar{c}$ with dimensions $[k]^{-2}$ and time dependence $\bar{c}= [{\cal H}f(\tau)D(\tau)]^2 c$, where $c$ is time independent. This time scaling is chosen to match the time scaling of one-loop SPT contributions.
In Sec.~\ref{sec:trispectrum}, we perform an analysis of these shapes that further reduces the operator basis to a linearly independent set.

\subsection{Construction of the EFT Kernels}\label{sec:kernels}
Now that we have constructed the stress tensor, what remains is to derive the form of $\tilde{\delta}_3$ which appears in EFT counterterm diagrams. Plugging in the ansatz from Eqs.~(\ref{eq:ansatzd}-\ref{eq:ansatzw}) and keeping all $\mathcal{O}(\varepsilon)$ terms, we find that the diagonalized equations of motion Eqs.~(\ref{diagdelta}-\ref{diagtheta}) and equation for the vorticity Eq.~(\ref{omegaeq2}) read
\begin{align}
-\frac{1}{2} (n+1 )(2n + 7)  \tilde{\delta}_n & =  \left(n+\frac{5}{2}\right)S_{\alpha,n} + S_{\beta,n}  -\partial_i \left(\frac{\partial_j \uptau^{ij}}{1+\delta_l}\right)_n \\
\frac{1}{2} (n+1 )(2n + 7) \tilde{\theta}_n  & =-\frac{3}{2}S_{\alpha,n} - \left(n+2\right) S_{\beta,n} +\left(n+2\right) \partial_i \left(\frac{\partial_j \uptau^{ij}}{1+\delta_l}\right)_n\\
 \left(n+\frac{5}{2}\right) \tilde{\omega}^i_n & = \epsilon^{ijk} \partial_j \left(\frac{\partial_b \uptau^{kb}}{1+\delta_l}\right)_n
\end{align}
where the time dependence has dropped out and where $n$ tracks the overall perturbative order.
Thus, for our purposes at order $n=3$ we find
\begin{align*}
\tilde{\delta}_3 &= -\frac{1}{26} \partial_i \partial_j \uptau^{ij}_3 +\frac{1}{26} \partial_i (\delta_1 \partial_j \uptau^{ij}_2) +\frac{1}{26} \partial_i ((\delta_2 - \delta_1^2) \partial_j \uptau^{ij}_1) + \frac{11}{234} \alpha^i_\omega \delta_1 \epsilon_{ijk} \partial^j \partial_b \uptau^{kb}_2  \\
&+\frac{1}{117} \beta^i_\omega \theta_1 \epsilon_{ijk} \partial^j \partial_b \uptau^{kb}_2 +\frac{1}{26} (\theta_2 \tilde{\theta}_1 + \tilde{\theta}_2 \theta_1)\beta  + \frac{11}{52} (\theta_2 \tilde{\delta}_1 + \tilde{\theta}_2 \delta_1 + \theta_1 \tilde{\delta}_2 + \tilde{\theta}_1 \delta_2)\alpha  \numberthis
\end{align*}
where for compactness we have left the arguments of the SPT mode-coupling kernels and the convolution integrals implicit. Terms involving the Levi-Civita symbol come from the vorticity kernel which is sourced by $\uptau_2^{ij}$. Note that this quantity is directly related to the kernel $\widetilde{F}_3$ through Eq.~\eqref{eq:kernels}, and will have to be symmetrized when written explicitly in terms of wavenumbers. 

\section{SPT Renormalization and the Minimal Basis of EFT Operators}
\label{sec:trispectrum}

Having specified the most general stress tensor through NNLO, \emph{i.e.} $\order(\delta_1^3)$, we may now compute the trispectrum at one loop, examining both the UV and finite contributions. We note that individual one-loop diagrams exhibit IR divergences as well, but they cancel once all the diagrams are summed over (see \emph{e.g.} \cite{2014JCAP...07..056C}). In the following, we refer extensively to the EFT kernels ${\widetilde F}_n, {\widetilde G}_n, {\widetilde G}^{\omega}_n$ defined in Eq.~\eqref{eq:kernels}, whose explicit forms are collected in Appendix~\ref{app:kernels}. We report the kernels with a minimal set of EFT coefficients whose corresponding shapes in Fourier space form a complete basis that spans all possible EFT operators at leading order in $\varepsilon$. Our convention for the diagrammatic representation for the various perturbative contributions follows that of~\cite{Scoccimarro:1995if}.

\subsection{UV Contributions}
\label{sec:UV}
Let us first focus on the contributions from regions of integration where the loop wavenumber $q$ is much larger than the external wavenumbers. In this limit, the symmetrized SPT kernels in Eq.~\eqref{eq:kernels} exhibit a universal scaling behavior $F_n \sim k^2/q^2$, where $k$ represents the characteristic scale of the fixed external wavenumbers. In the present analysis, we will focus on the renormalization of the diagrams that scale in the UV as $\sim k^2/q^2$, neglecting the renormalization of subleading UV contributions that scale with higher powers of the loop wavenumber. The diagrams with this leading-order UV behavior are the ones for which the loop integral includes only one mode-coupling vertex (not counting $F_1$), such as $P_{31}$ in \Fig{fig:powerspec}, which depends only on $F_3$. In the UV limit, the kernels $F_n$ for $n \geq 3$ have the form
\begin{align}\label{eq:UVscaling}
 \lim_{q \gg k_i} F_n(\vec{q},-\vec{q},\vec{k}_1, \dots ,\vec{k}_{n-2}) &
 \equiv \sum_{r=1}^\infty {F_n^{(2r)}(\vec{k}_1, \dots ,\vec{k}_{n-2}) \over q^{2r}} = {F_n^{(2)} (\vec{k}_1, \dots ,\vec{k}_{n-2}) \over q^2}
 + \dots  \,,
\end{align}
where the ellipsis denotes subleading terms. Loop integrals are always accompanied by a loop propagator in the form of the linear matter power spectrum, $P_L(q)$---
thus, the leading UV contribution is proportional to $\sigma^2 \equiv 1/3 \int \dbar^3 q { P_L(q) \over q^2}$, 
which represents the dependence of the diagrams on highly nonlinear modes, $q \gg k_i$. This is unphysical (albeit finite due to regularization from $P_L(q)$ in $\Lambda$CDM) and should be accounted for through renormalization. The contributions from the EFT of LSS cancel this UV dependence via the counterterms in Eq.~\eqref{eq:stresstensor}. Since the SPT kernels have a highly nontrivial functional form even in the UV, this cancellation is a stringent consistency check of the theory. In particular, the lower-point correlation functions and the cancellation of their divergences propagate through the calculation of the higher-point correlation functions. Here, we systematically carry out the renormalization from the power spectrum up to the trispectrum to explicitly show the self-consistency of our calculation.

\subsubsection{Power Spectrum}
\begin{figure*}[htb]
\begin{center}
\includegraphics[width=\textwidth]{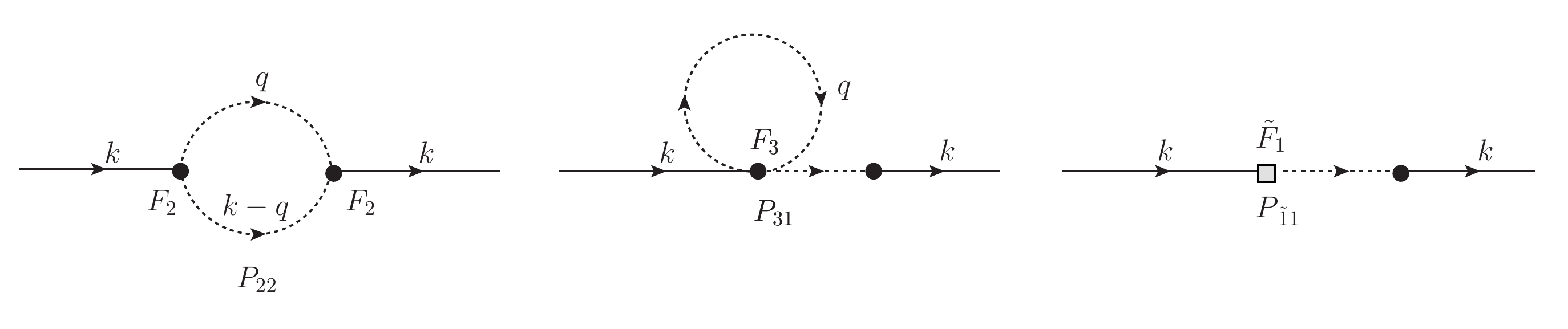}
\caption{The one-loop SPT and EFT counterterm diagrams for the power spectrum. Circular (square) vertices denote insertions of the SPT (EFT) kernels.}
\label{fig:powerspec}
\end{center}
\end{figure*}
The one-loop diagrams for the power spectrum are shown in Fig.~\ref{fig:powerspec}.  
From the UV scaling discussed in Sec.~\ref{sec:UV}, we see that the UV contribution of the $P_{22}$ diagram is subleading, scaling as $ 1/q^4$, while that of the $P_{31}$ diagram scales as $ 1/q^2$. The counterterm $P_{{\tilde 1}1}$ is proportional to the $\widetilde{F}_1$ kernel given in Appendix~\ref{app:kernels}. The sum of the two amplitudes is given by
\begin{align} 
P_{{\tilde 1}1} + P_{31}^{\rm UV}   &= - {c_s k^2 \over 9} P_L(k) + \frac{3!}{2!} \int \dbar^{\,3}q \, {F^{(2)}_3(\vec{k})  \over q^2} P_L(q) P_L(k) \nonumber  \\
&=  - {c_s k^2 \over 9} P_L(k) - \frac{61 k^2 \sigma^2}{210} P_L(k)
.\end{align}
Thus, the one-loop power spectrum is renormalized with
\beq \label{eq:cs}
c_s = - \frac{183}{70} \sigma^2 + c_s^{\rm ren}\,,
\eeq
where $c_s^{\rm ren}$ is the finite (``renormalized'') piece of the counterterm, which is independent of the form of the $\sigma^2$ piece which accounts for the purely UV behavior. At the relevant scales, the finite part of $c_s$ is determined from measurements, for instance of power spectrum data.

\subsubsection{Bispectrum}
\begin{figure*}[htb]
\begin{center}
\includegraphics[width=\textwidth]{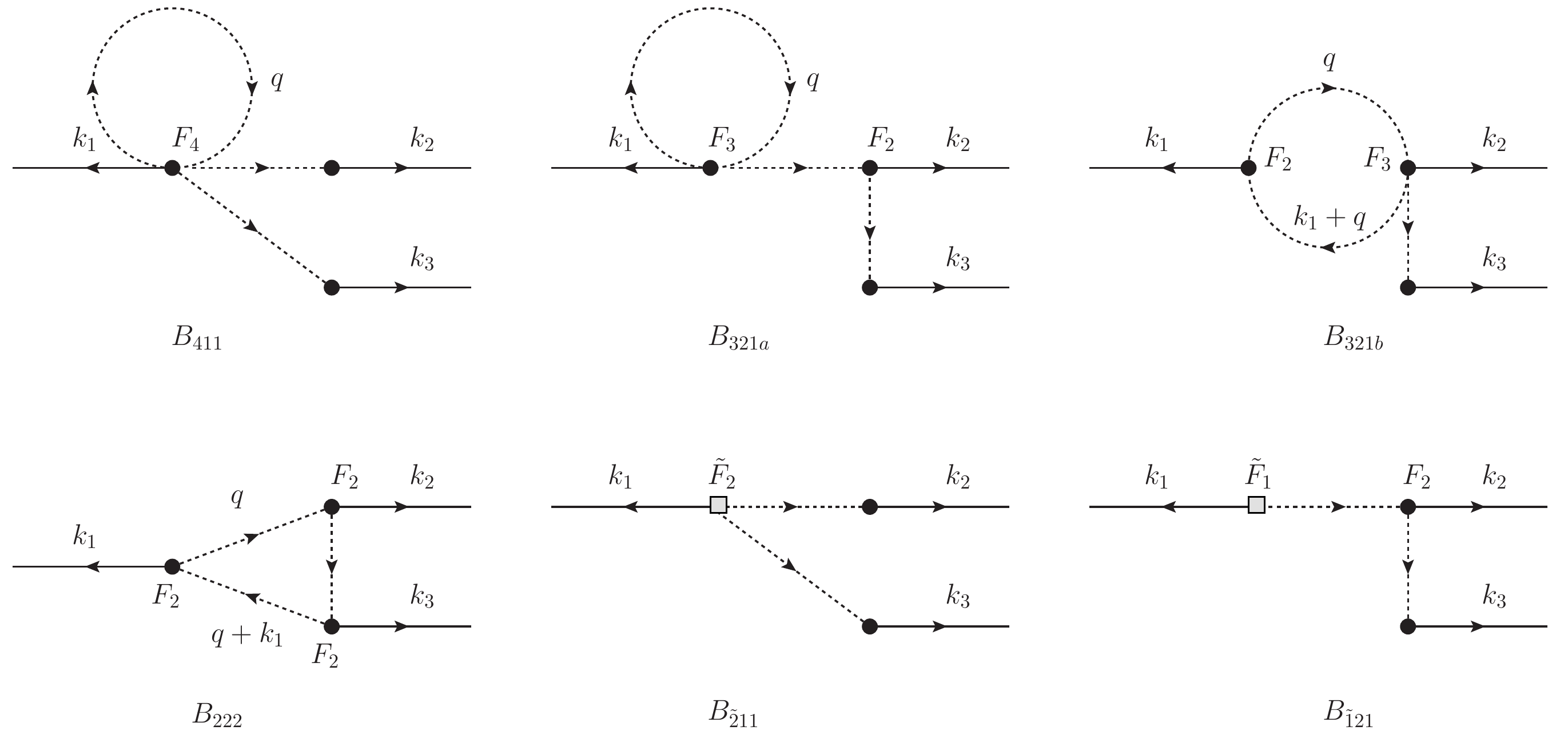}
\caption{The one-loop SPT and EFT counterterm diagrams for the bispectrum. Circular (square) vertices denote insertions of the SPT (EFT) kernels.}
\label{fig:bispectrum}
\end{center}
\end{figure*}
The one-loop diagrams for the bispectrum are shown in Fig.~\ref{fig:bispectrum}. From the UV scaling discussed in \Sec{sec:UV}, we see that the UV contributions from the $B_{222}$ and $B_{321b}$ diagrams scale as $1/q^6$ and $1/q^4$, respectively, and are thus subleading.
Additionally, notice that the UV dependence of the $B_{321a}$ diagram will be exactly the same as that of the $P_{31}$ diagram of the power spectrum, and hence the renormalization of the former follows from that of the latter (we refer the reader to \cite{2015JCAP...05..007B, 2014arXiv1406.4143A} for further details). The remaining diagram is $B_{411}$, and its UV contribution is given by
\begin{align}
B_{411}^{\rm UV} &= \frac{4!}{2!}  \int \dbar^{\,3}q \, {F_4^{(2)}( \vec{k}_2, \vec{k}_3) \over q^2} P_L(q) P_L(k_2) P_L(k_3) + \text{2 permutations} \nonumber \\
&=-\frac{\sigma^2}{226380 k_2^2 k_3^2} \ \bigg[12409 k_1^6+\left( 12024
   \left(k_2^2-k_3^2\right)^2+20085 k_1^4\right)
   \left(k_2^2+k_3^2\right) \nonumber \\
   &\quad+k_1^2 \left(76684
   k_2^2 k_3^2 -44518 (k_2^4+ k_3^4)\right)\bigg]P_L(k_2) P_L(k_3) + \text{2 permutations}\, ,
 \numberthis  
 \end{align}
where we have included contributions from permutations of the external wavenumbers. 
The contribution from the counterterm $B_{{\tilde 2}11}$ is given by
\begin{align}
B_{{\tilde 2}11} &= \Bigg[ \frac{-12 k_1^6-32 k_1^4 \left(k_2^2+k_3^2\right)+k_1^2 \left(23 k_2^4-74 k_2^2 k_3^2+23 k_3^4\right)+21 \left(k_2^2-k_3^2\right)^2 \left(k_2^2+k_3^2\right)}{1386 k_2^2 k_3^2}  c_s \nonumber \\
&\quad - {4k_1^2 \over 33 } c_1 -\frac{k_1^4 \left(k_2^2+k_3^2\right)-2 k_1^2 \left(k_2^2-k_3^2\right)^2+\left(k_2^2-k_3^2\right)^2 \left(k_2^2+k_3^2\right)}{66 k_2^2 k_3^2} c_2  \nonumber  \\
&\quad -\frac{\left(k_2^2-k_1^2-k_3^2\right) \left(k_1^2+k_2^2-k_3^2\right) \left(k_2^2+k_3^2-k_1^2\right)}{66 k_2^2 k_3^2} c_3 \Bigg]P_L(k_2) P_L(k_3) + \text{2 permutations}. \,
\end{align}
The $\widetilde{F}_2$ kernel and the definition of the coefficients $c_{1,2,3}$ in terms of those appearing in Eq.~\eqref{eq:stresstensor} is given in Appendix~\ref{app:kernels}. The $B_{411}$ diagram is renormalized with $c_s$ as given in Eq.~\eqref{eq:cs}, and
 \begin{align}
 c_1 = \frac{6077}{6860} \sigma^2 + c_1^{\rm ren}, \quad\quad
  c_2 = -\frac{979}{245}\sigma^2 + c_2^{\rm ren},\quad\quad
  c_3 = -\frac{1457}{686} \sigma^2 + c_3^{\rm ren},
  \end{align} 
where the finite renormalized coefficients $c_{1,2,3}^{\rm ren}$ are again determined from measurements, e.g., of bispectrum data.
Note that the requirements on $c_s$ from the renormalization of the power spectrum and bispectrum are a consistency check of the calculation.

 \subsubsection{Trispectrum}
 \label{sec:tri}
   \begin{figure*}[h!]
\begin{center}
\includegraphics[width=\textwidth]{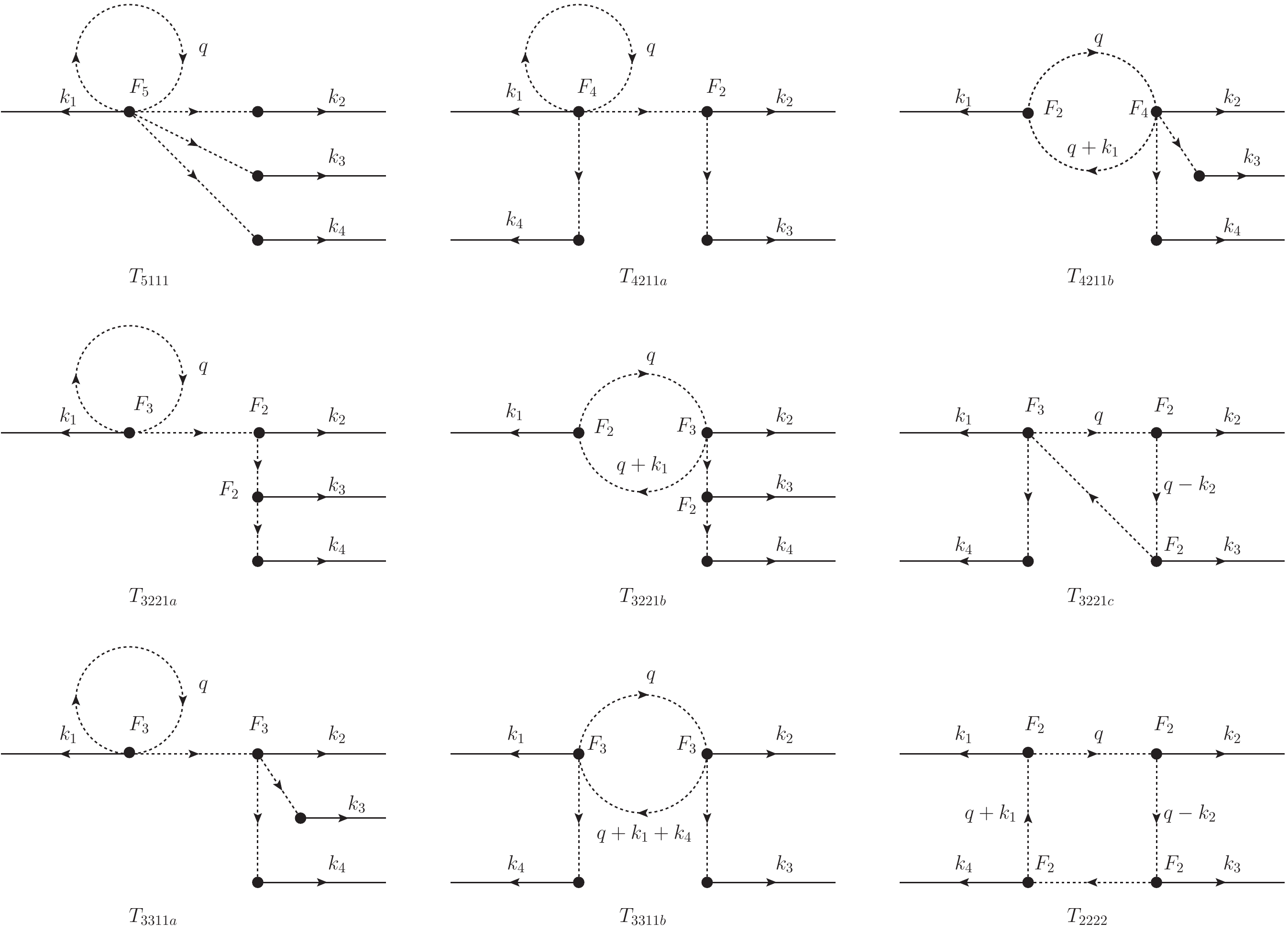}
\caption{One-loop SPT diagrams for the trispectrum. }
\label{fig:trispectrum}
\end{center}
\end{figure*}
   \begin{figure*}[h!]
\begin{center}
\includegraphics[width=0.8\textwidth]{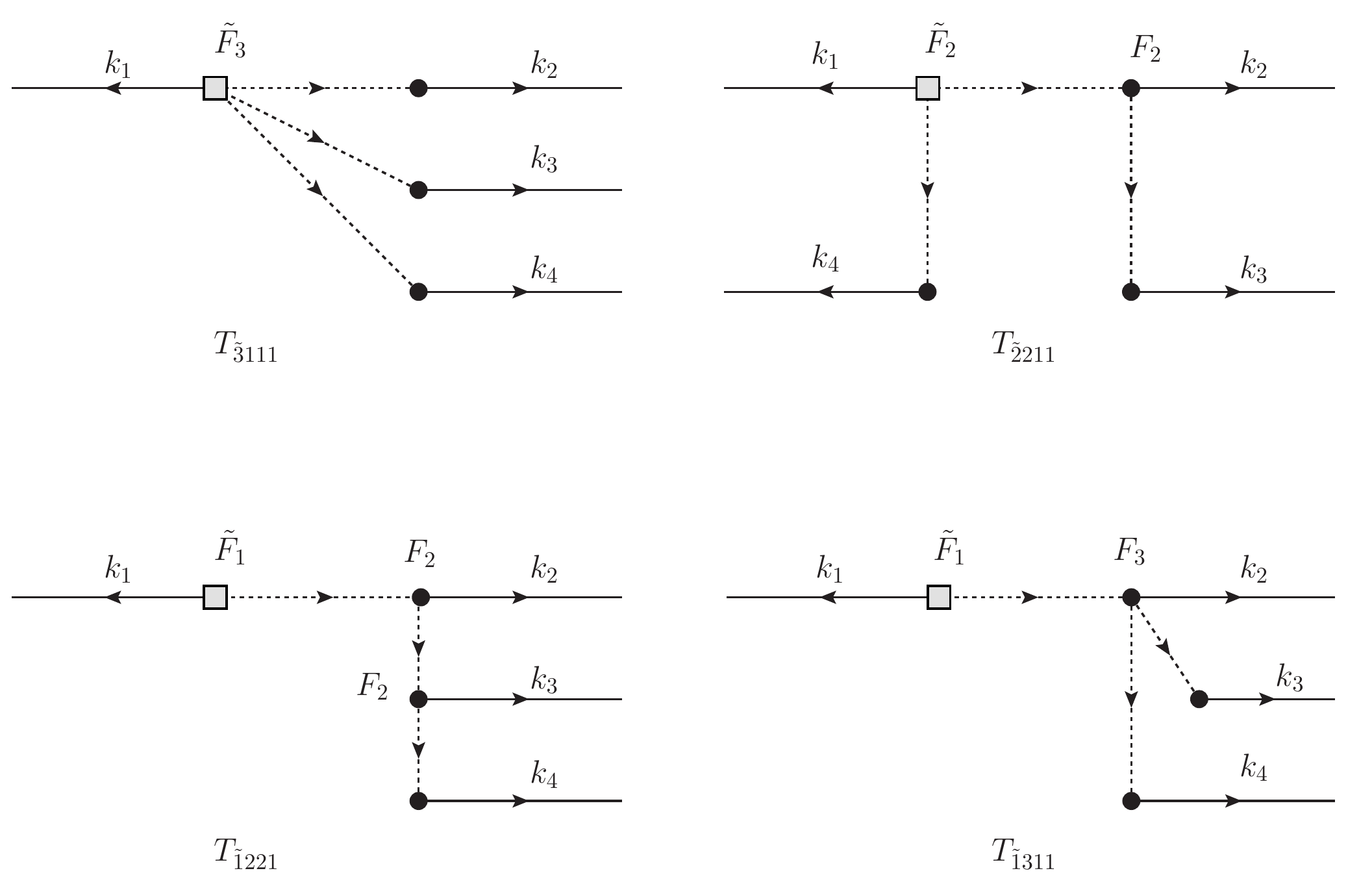}
\caption{Counterterm diagrams for the one-loop trispectrum. }
\label{fig:trispectrumEFT}
\end{center}
\end{figure*}

The one-loop SPT and EFT counterterm contributions to the trispectrum are shown in Figs.~\ref{fig:trispectrum} and~\ref{fig:trispectrumEFT}, respectively. As before, the diagrams involving multiple $F_n$ kernels, for $n \geq 2$, have subleading UV behavior since each kernel scales as $1/q^2$. In this case, the leading order UV diagrams are $T_{5111}$, $T_{4211a}$, $T_{3221a}$, and $T_{3311a}$. The renormalization of $T_{4211a}$ follows from that of $B_{411}$, while the renormalization of $T_{3221a}$ and $T_{3311a}$ follows from that of $P_{31}$. The diagram with leading-order UV dependence is $T_{5111}$, and has the form
\beq T_{5111}^{\rm UV} =\frac{5!}{2!} \int \dbar^{\,3}q \, { F_5^{(2)}(\vec{k}_2,\vec{k}_3,\vec{k}_4) \over q^2}  P_L(q) P_L(k_2) P_L(k_3) P_L(k_4) + \text{3 permutations.} 
\eeq 
The full expression for $F_5^{(2)}(\vec{k}_2,\vec{k}_3,\vec{k}_4)$ is too long to include here. The UV behavior of this diagram is renormalized by the counterterm diagram $T_{{\tilde 3}111}$.
A straightforward calculation shows that the renormalization is consistent with the values for $c_s$ and $c_{1,2,3}$ fixed by the power spectrum and bispectrum, and involves eight new independent operators, whose coefficients are fixed to
\begin{equation}\label{eq:triUV}
\begin{split}
d_1 &= \frac{2658583}{4753980} \sigma^2+d_1^\text{ren},\\
d_3 &= -\frac{167627}{113190}\sigma^2+d_3^\text{ren},\\
d_5 &= \frac{137947}{45276}\sigma^2+d_5^\text{ren},\\
c_4 &= -\frac{934103}{75460}\sigma^2+c_4^\text{ren},
\end{split}
\quad\quad
\begin{split}
d_2 &= -\frac{527117}{75460}\sigma^2+d_2^\text{ren},\\
d_4 &= -\frac{33053}{1584660}\sigma^2+d_4^\text{ren},\\
d_6 &= \frac{72911}{37730}\sigma^2+d_6^\text{ren},\\ 
c_5 &= \frac{22147}{12936}\sigma^2+c_5^\text{ren}.
\end{split}
\end{equation}
The $\widetilde{F}_3$ kernel and the definition of the coefficients in terms of those appearing in Eq.~\eqref{eq:stresstensor} are given in Appendix~\ref{app:kernels}. In particular, of the eight new EFT operators appearing at this order, those corresponding to the coefficients $d_{1...6}$ have been chosen as the shapes $E_{1...6}$ listed in Eq.~\eqref{eq:operators}. The remaining two operators, corresponding to the $c_{4,5}$ coefficients, arise from the propagation of NLO operators. These are shapes proportional to $\theta\theta$ that are completely degenerate with operators proportional to $\delta\delta$ at NLO, but are independent at NNLO, and are needed for a consistent renormalization of the trispectrum.

Trispectrum data is currently not available, but one can in principle measure the finite parts of some linear combination of these coefficients from the two-loop power spectrum and the one-loop covariance. Consistency of the measured parameters across different observables would provide a strong check on the EFT of LSS.

\subsection{Finite Contributions}
\label{sec:finite}

We will now illustrate the dependence of the trispectrum on external wavevectors by looking at the finite contributions. Since measuring the primordial trispectrum from inflation \cite{2007JCAP...01..027S,2007JCAP...01..008S,2006PhRvD..74l3519B} is one of the main motivations for predicting the form of this observable, we focus on configurations where the primordial trispectrum is expected to be largest, following Ref. \cite{2010PhRvD..82b3520R}. For ease of comparison, we employ the labels adopted there, and note that the variables $\epsilon$, $\alpha$, $\beta$, $\delta$, $\theta$ and $\gamma$ defined in this section are \emph{only} for parametrizing quadrilateral configurations and should \emph{not} be confused with their usage in previous sections. For our numerical analysis, we use a linear power spectrum from CAMB~\cite{Lewis:1999bs}, with the following cosmological parameters: $\Omega_m=0.286$, $\Omega_b=0.047$, h$=0.7$, $n_s=0.96$, and $\sigma_8=0.82$.

A generic trispectrum configuration $T(\vec{k}_1,\vec{k}_2,\vec{k}_3,\vec{k}_4)$ is specified by six numbers, which for instance can be taken as the magnitudes of the four external wavenumbers and the two diagonals of the (non-planar) quadrilateral. Without loss of generality, let us consider vectors $\vec{k}_1$ and $\vec{k}_2$ to be on the same plane, ${\cal P}_{12}$, while the vectors $\vec{k}_3$ and $\vec{k}_4$ to be on another plane, ${\cal P}_{34}$. We then adopt the following parametrization:
\begin{align}
\label{eq:quadri}
|\vec{k}_1|&=\frac{k}{2}(1+\alpha+\beta),\quad |\vec{k}_2|=\frac{k}{2}(1-\alpha+\beta), \quad |\vec{K}| = k(1-\beta) = \epsilon k (1-\delta) ,  \nonumber \\ 
|\vec{k}_3|&=\frac{\epsilon k}{2}(1+\gamma+\delta), \quad |\vec{k}_4|=\frac{\epsilon k}{2}(1-\gamma+\delta), \quad \cos \theta = 
\vec{k_1} \cdot \vec{k}_3/|\vec{k}_1||\vec{k}_3| \,,
\end{align}
where $\vec{K} = \vec{k}_1 + \vec{k}_2 = \vec{k}_3 + \vec{k}_4$, thus relating $\delta$ and $\beta$. Moreover, instead of employing the second diagonal, we have introduced the angle $\theta$ between the planes ${\cal P}_{12}$ and ${\cal P}_{34}$.

\begin{figure}[htb]
\begin{center}
\includegraphics[width=\textwidth]{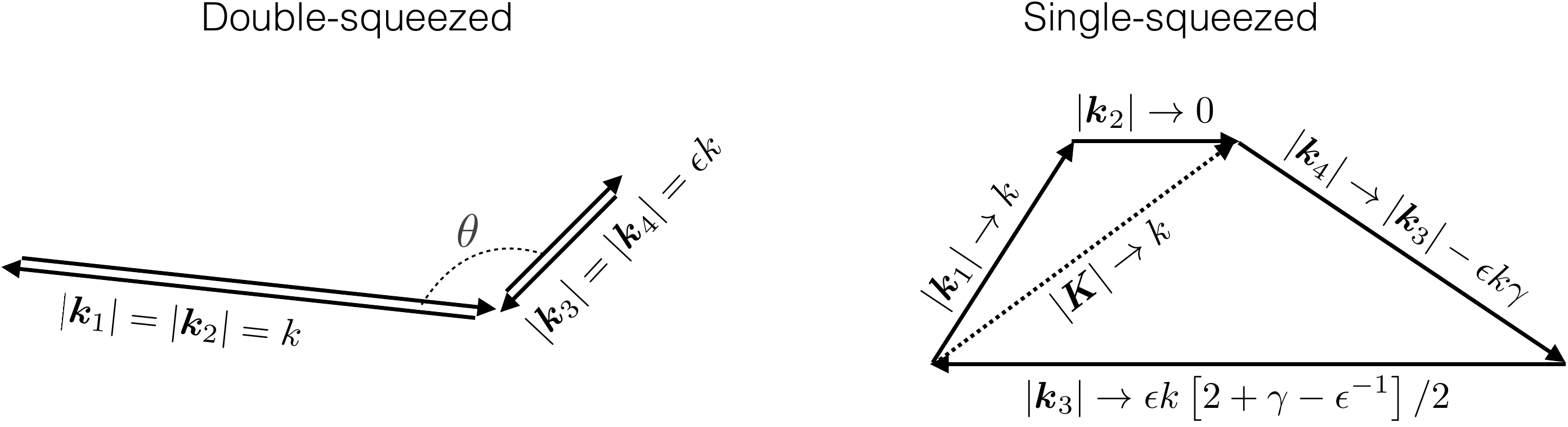}
\caption{Double- and single-squeezed trispectrum configurations analyzed in \Sec{sec:finite}. We adopt the same parametrization as in Ref. \cite{2010PhRvD..82b3520R}.}
\label{fig:configs}
\end{center}
\end{figure}

We will consider two configurations: the double-squeezed and single-squeezed limits illustrated in Fig.~\ref{fig:configs}, which correspond to peaks of primordial non-Gaussianity signals \cite{2010PhRvD..82b3520R}, for local models of the primordial trispectrum. Additionally, the equilateral model corresponds to the case $\epsilon=1$ and $\theta=\pi$ of the double-squeezed configuration. 
In the double-squeezed limit both triangles forming the quadrilateral are squeezed, resulting in a configuration with no triangle. This limit is given by $\alpha, \gamma \to 0$ and $\beta, \delta \to 1$, implying $|\vec{K}| \to 0$.
In the single-squeezed limit, only one of the two triangles forming the quadrilateral is squeezed
and the configuration corresponds, {\it e.g.}, to $\beta\to 0$ and $\alpha\to 1$, which implies $|\vec{k}_2|\to 0$, $|\vec{K}| \to k$ and $\delta \to 1-1/\epsilon$. The resulting triangle is uniquely determined by $k$, $\epsilon$ and $\gamma$, where the last two can take values $\epsilon\geq 1$ and $-1/\epsilon \leq \gamma \leq 1/\epsilon$.
\begin{figure}[H]
\begin{center}
\begin{subfigure}[b]{0.48\textwidth} 
\includegraphics[width=\textwidth]{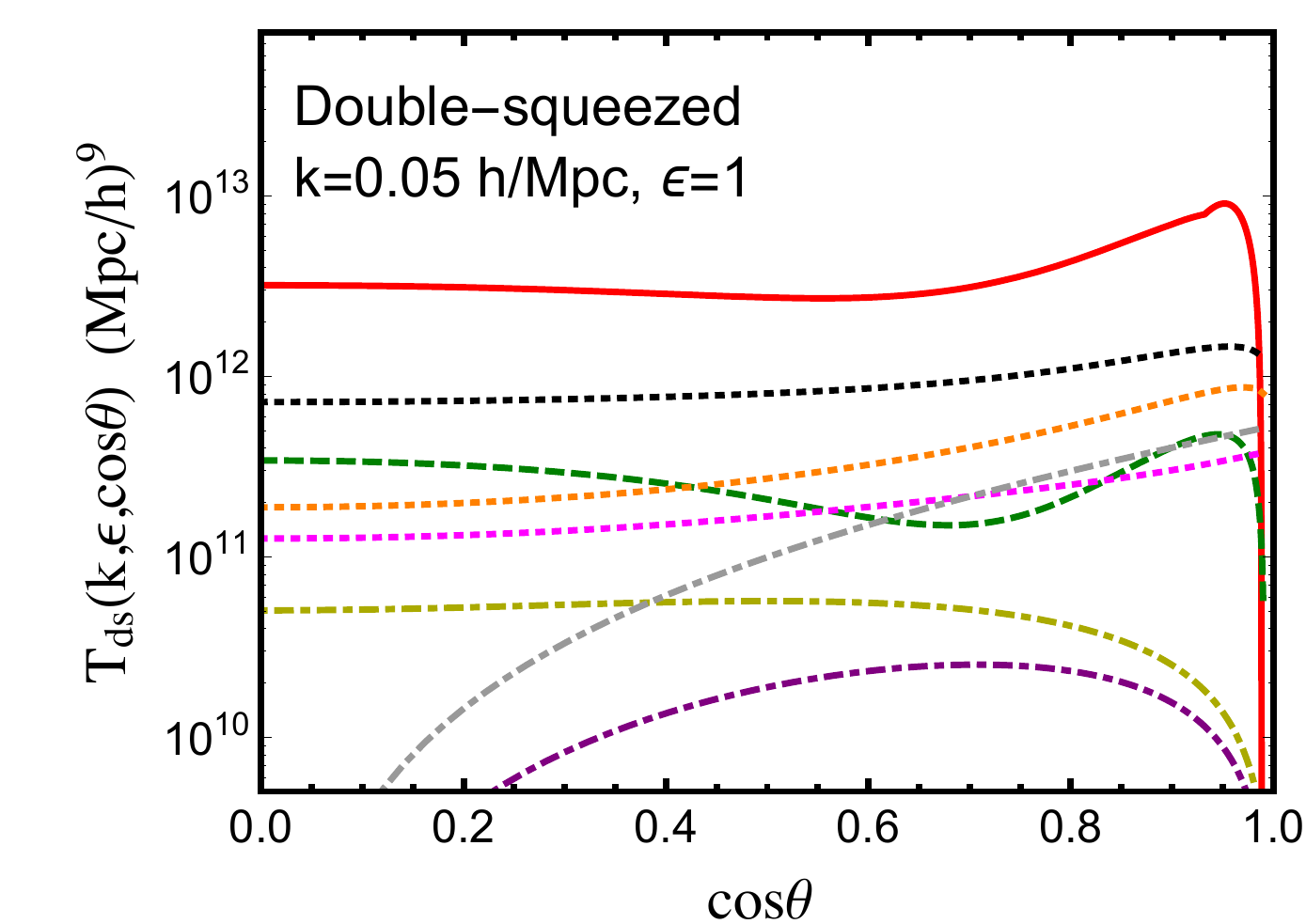}
\caption{}\label{a}
\end{subfigure}
\quad
\begin{subfigure}[b]{0.48\textwidth} 
\includegraphics[width=\textwidth]{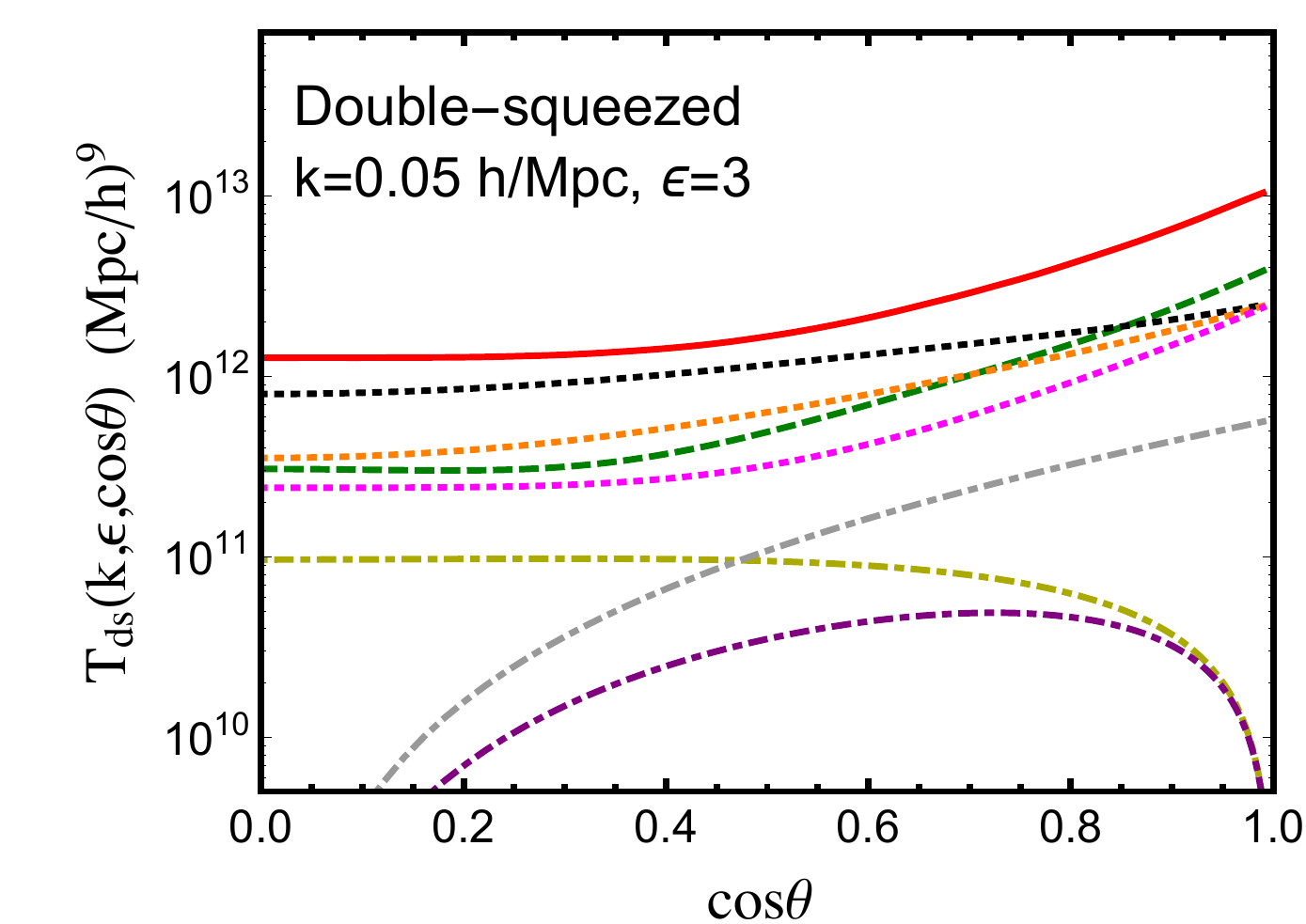}
\caption{}\label{b}
\end{subfigure}
\begin{subfigure}[b]{0.9\textwidth} 
\includegraphics[width=\textwidth]{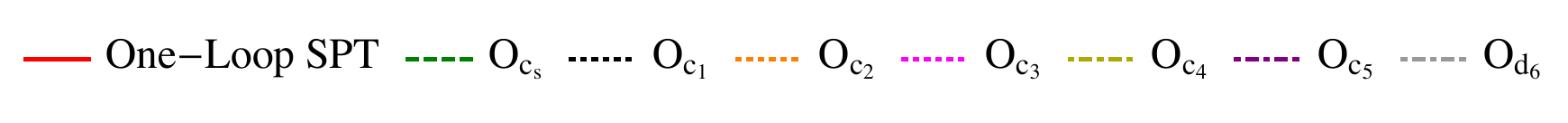}
\end{subfigure}
\begin{subfigure}[b]{0.48\textwidth} 
\includegraphics[width=\textwidth]{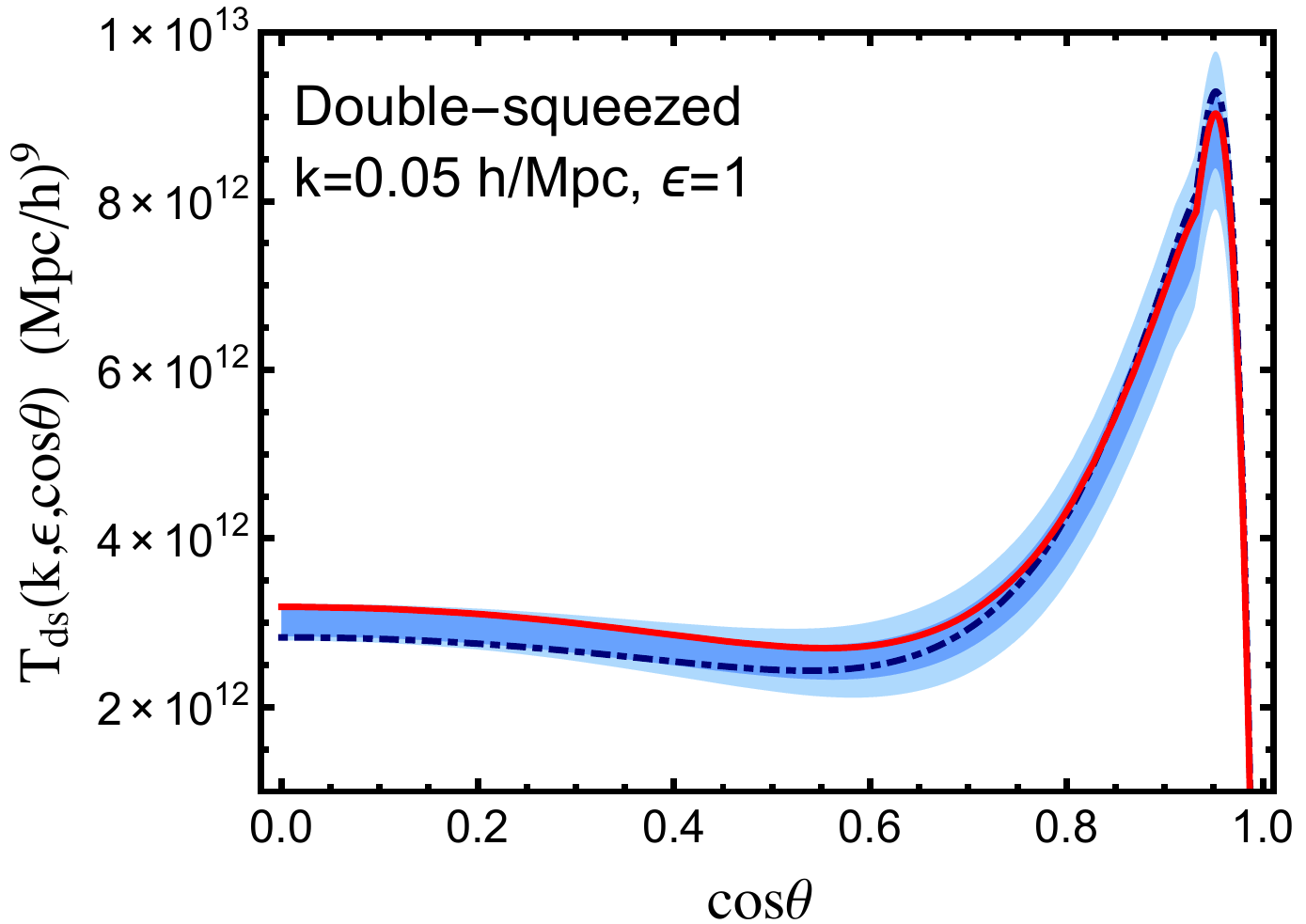}
\caption{}\label{c}
\end{subfigure}
\quad
\begin{subfigure}[b]{0.48\textwidth} 
\includegraphics[width=\textwidth]{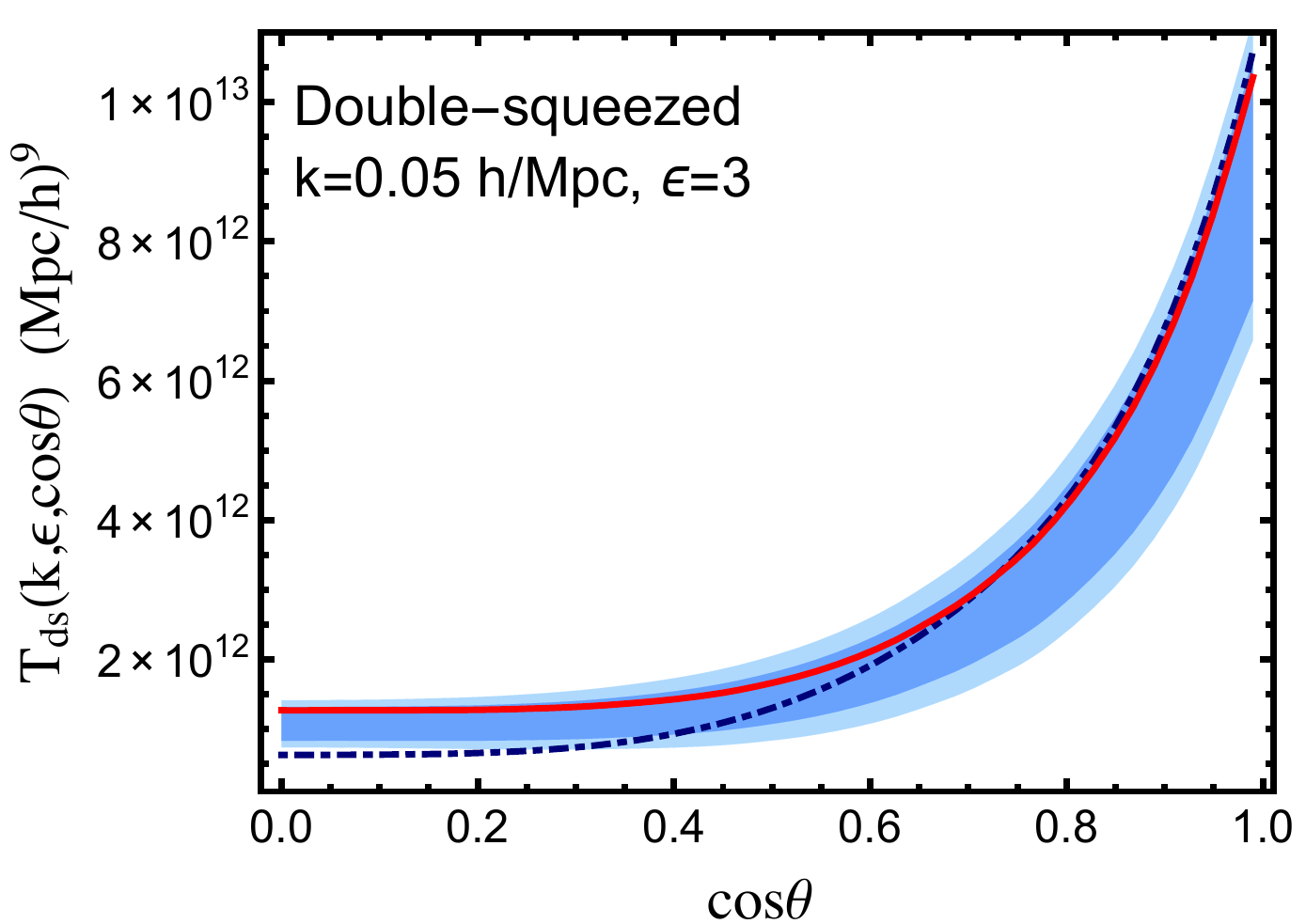}
\caption{}\label{d}
\end{subfigure}
\begin{subfigure}[b]{0.9\textwidth} 
\includegraphics[width=\textwidth]{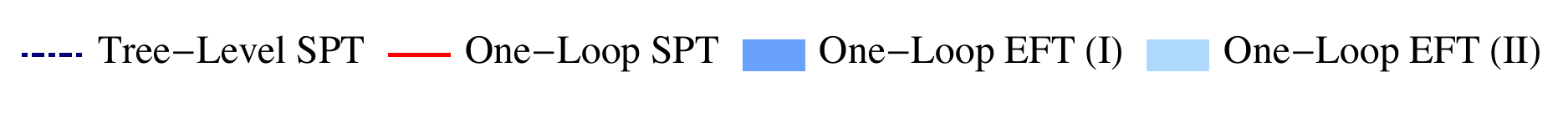}
\end{subfigure}
\caption{One-loop SPT (solid red) and EFT (dashed, dotted, and dot-dashed) contributions to the trispectrum in the double-squeezed configuration (see \Eq{eq:quadri} and below). The EFT operators $O_x$ correspond to the contributions from the sum of the $T_{\tilde{3}111}$, $T_{\tilde{2}211}$, $T_{\tilde{1}221}$, and $T_{\tilde{1}311}$ diagrams proportional to the coefficient $x$. These contributions have been multiplied by $-10^2$, except for the operators proportional to $c_4$ and $c_5$ which are multiplied by $10^2$. Panels \ref{a} and \ref{b} show the case $k=0.05$ h/Mpc for $\epsilon=1,3$ as a function of the cosine of the angle between the two independent directions. Note that the plotted contributions are even functions of cosine.
Panels \ref{c} and \ref{d} show the predictions for SPT at tree-level (dot-dashed blue) and at one-loop (solid red), as well as for the one-loop EFT (blue bands). One-loop EFT (I) (dark blue band) corresponds to including only the previously measured coefficients $\{c_s, c_1, c_2, c_3\}$, and varying them around their best fit values by $50\%$. One-loop EFT (II) (light blue band) corresponds to including all the EFT coefficients and varying the unknown ones in the range $[-100,100] \ {\rm Mpc}^2/{\rm h}^2$.}
\label{fig:ds}
\end{center}
\end{figure}\newpage
\begin{figure*}[t]
\begin{center}
\begin{subfigure}[b]{0.48\textwidth} 
\includegraphics[width=\textwidth]{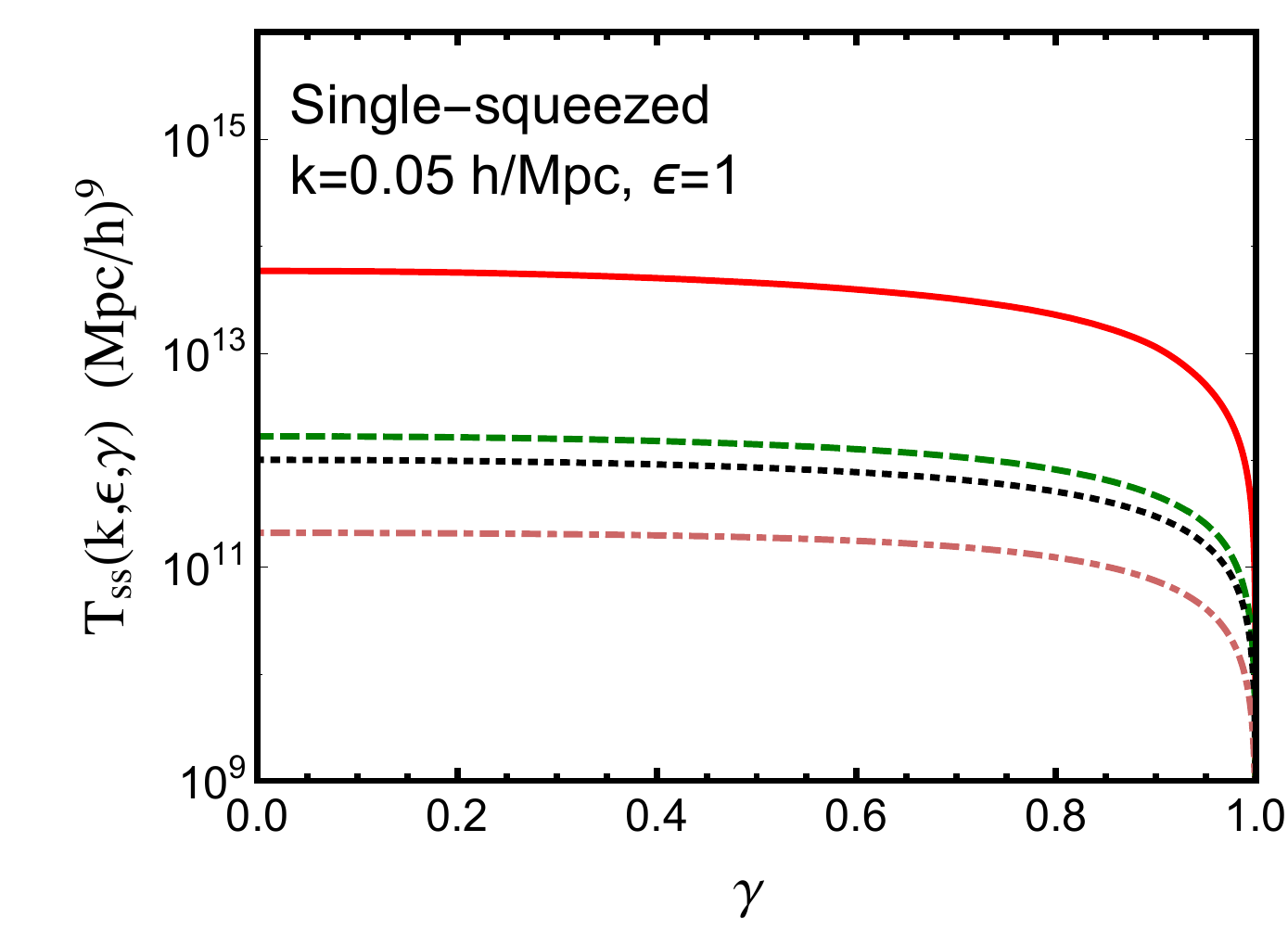}
\caption{}\label{a2}
\end{subfigure}
\quad
\begin{subfigure}[b]{0.48\textwidth} 
\includegraphics[width=\textwidth]{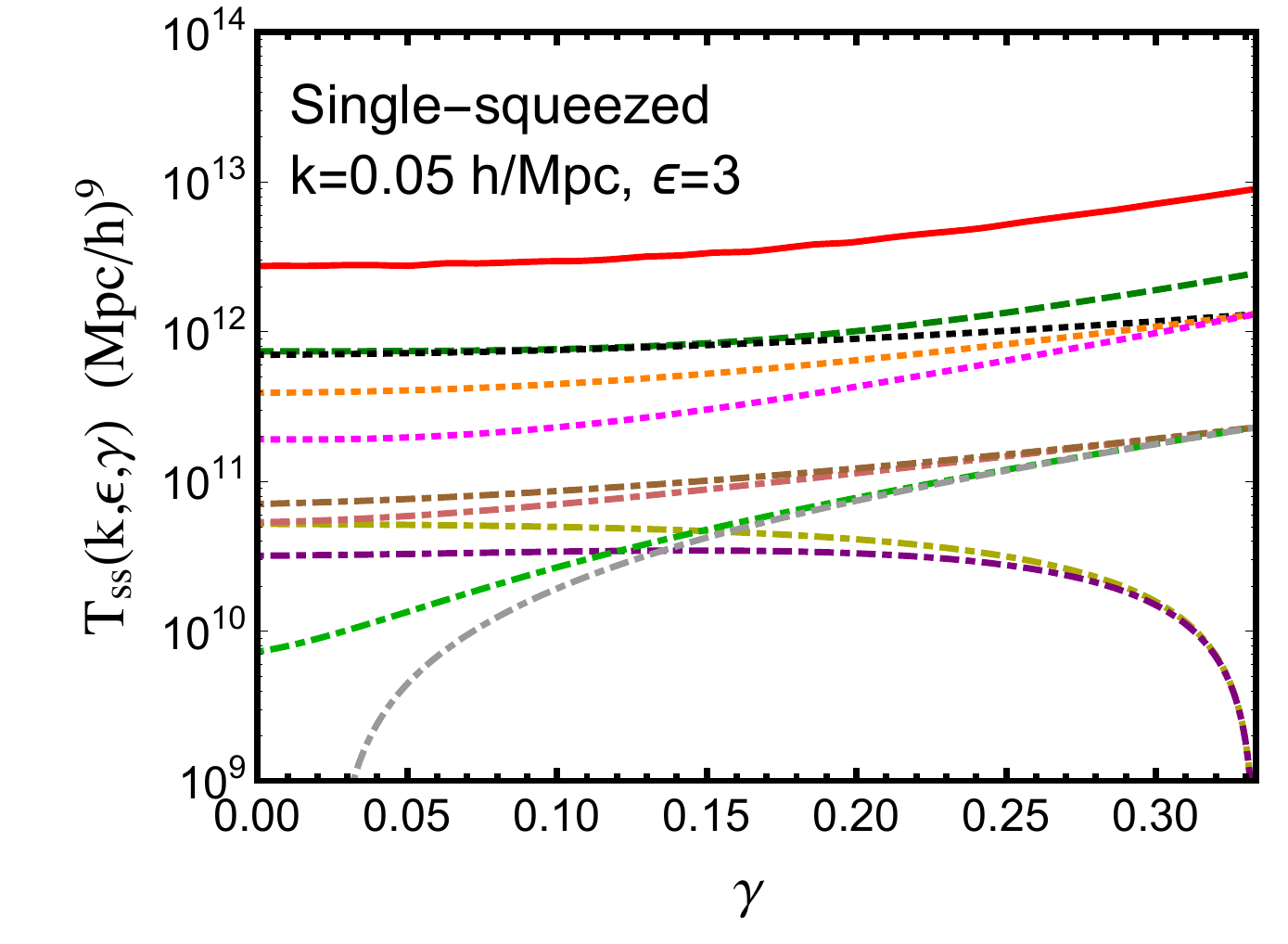}
\caption{}\label{b2}
\end{subfigure}
\begin{subfigure}[b]{0.7\textwidth} 
\includegraphics[width=\textwidth]{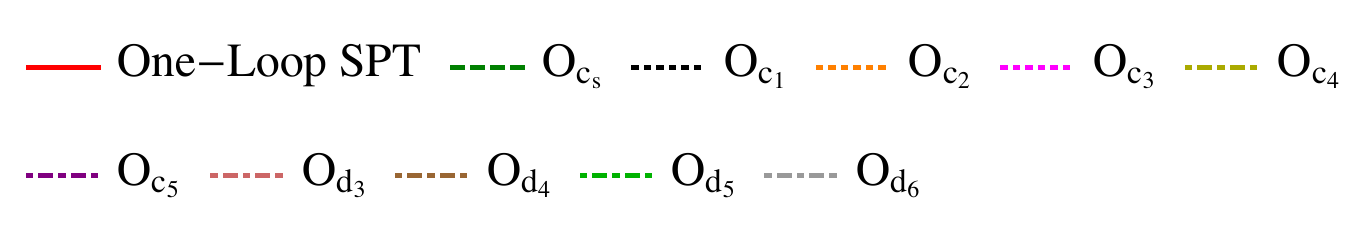}
\end{subfigure}

\begin{subfigure}[b]{0.48\textwidth} 
\includegraphics[width=\textwidth]{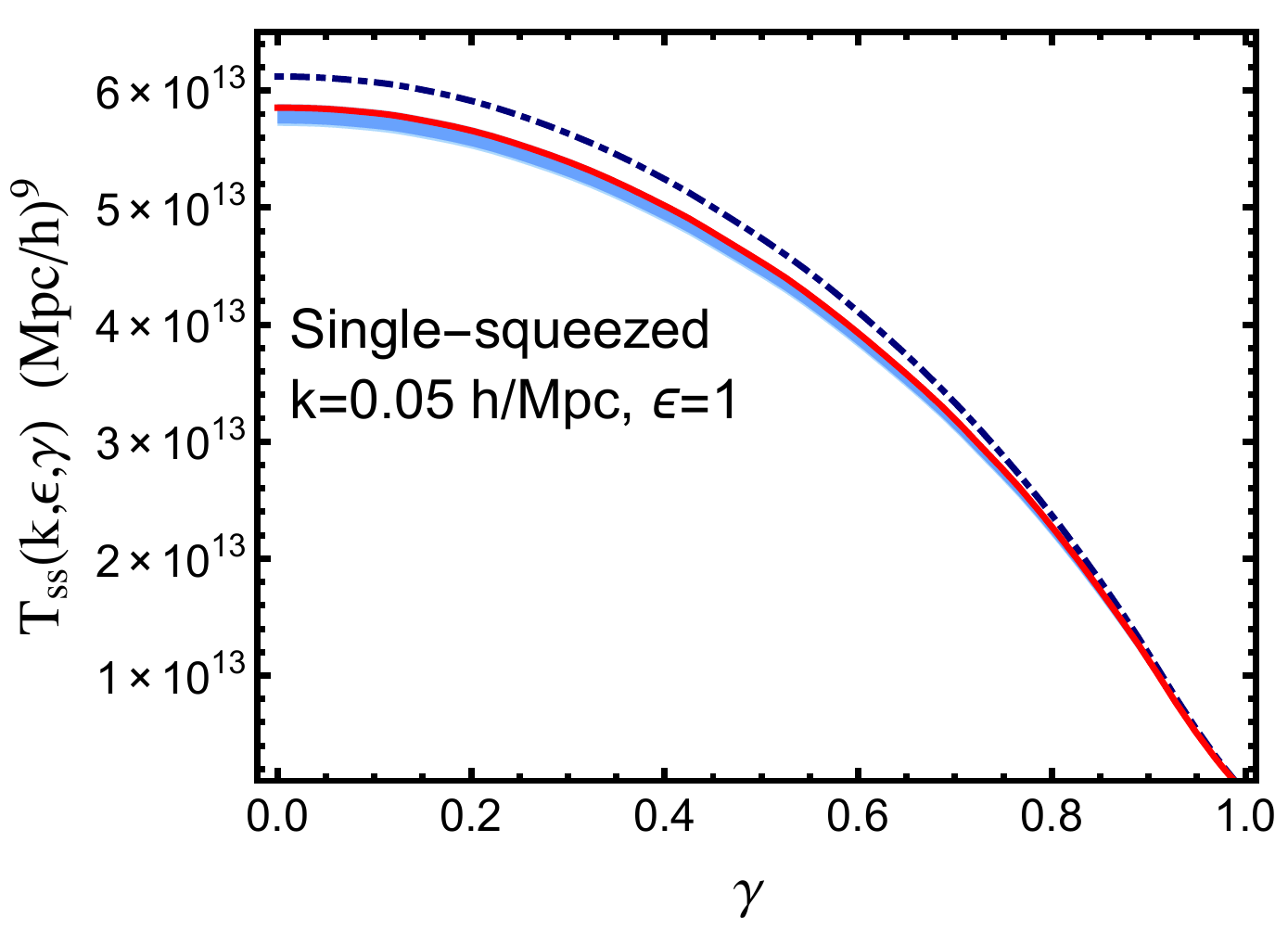}
\caption{}\label{c2}
\end{subfigure}
\quad
\begin{subfigure}[b]{0.48\textwidth} 
\includegraphics[width=\textwidth]{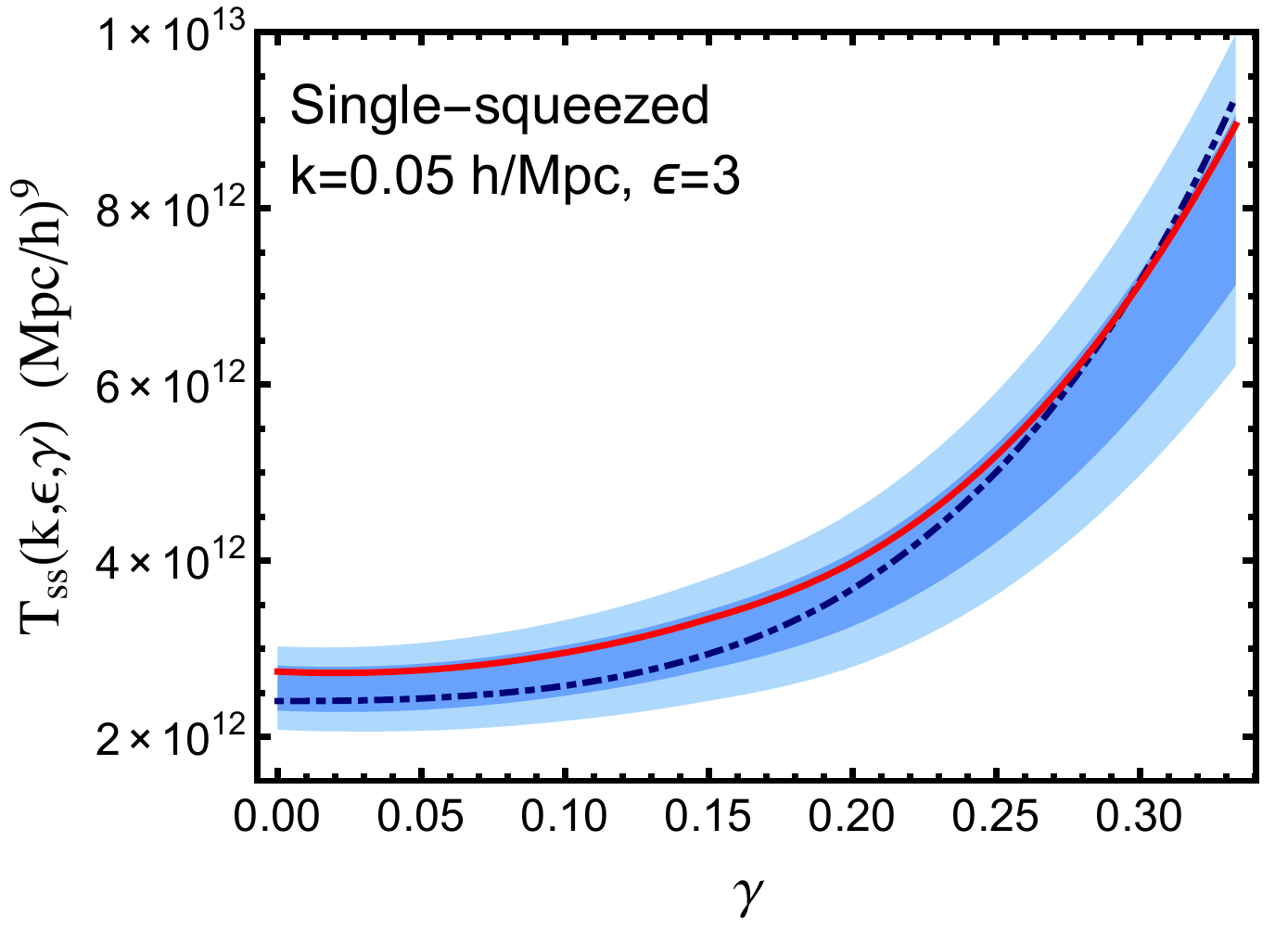}
\caption{}\label{d2}
\end{subfigure}
\begin{subfigure}[b]{0.9\textwidth} 
\includegraphics[width=\textwidth]{./Figures/legdsb.pdf}
\end{subfigure}
\caption{Same as \Fig{fig:ds}, but for the single-squeezed configuration. For the case $\epsilon=1$, $O_{c_1}=O_{c_2}=O_{c_3}$, thus we plot only $O_{c_1}$. Similarly $O_{d_3}=O_{d_4}=O_{d_5}=O_{d_6}$ and we show only $O_{d_3}$. Finally, $O_{c_4}$ and $O_{c_5}$ vanish. All the plotted contributions are even functions of $\gamma$.}\label{fig:ss}
\end{center}
\end{figure*}

In the upper panels of Figs.~\ref{fig:ds} and \ref{fig:ss}, we plot the contributions to the trispectrum from SPT, including one-loop diagrams, and the EFT counterterm diagrams. For the double-squeezed configuration we plot them as a function of $\cos\theta$ (between 0 and 1, since the functions are even in $\cos\theta$), and for the single-squeezed as a function of $\gamma$ (between 0 and $1/\epsilon$, since the functions are even in $\gamma$), with fixed values of $k$ and $\epsilon$. The values of $k$ and $\epsilon$ are chosen such that we are in the weakly nonlinear regime, where the one-loop corrections are small compared to the tree-level. As discussed in \Sec{sec:tri}, there are twelve independent operators at leading order in the 
derivative expansion of the stress-tensor,
for a generic trispectrum configuration. They reduce to seven operators in the double-squeezed limit and to ten in the single-squeezed limit, and can be chosen as those corresponding to the coefficients $\{c_s, c_1, c_2, c_3, c_4, c_5, d_6\}$ and $\{c_s, c_1, c_2, c_3, c_4, c_5, d_3, d_4, d_5,d_6\}$, respectively.

In the lower panels of Figs.~\ref{fig:ds} and \ref{fig:ss}, we show predictions for the trispectrum for SPT at tree-level (dot-dashed blue) and at one-loop (solid red), as well as for the EFT (blue bands). The dark band includes only the coefficients $\{c_s, c_1, c_2, c_3\}$ with the following best fit values taken from Ref.~\cite{2015JCAP...05..007B}:\footnote{See Ref.~\cite{2015arXiv151207630B} for the map between the operator basis used there and the basis used in this paper.}
$
{\bar c}_s=13.5, ~ {\bar c}_1=18.5,  ~{\bar c}_2=-41.1, ~ {\bar c}_3=62.4 \,,
$
in units of ${\rm Mpc}^2/{\rm h}^2$.
Note that these coefficients have the time dependence given below Eq. (\ref{eq:operators}). The band is formed by conservatively varying these values by $50\%$ simultaneously (an upper bound on the error). The light band includes all coefficients, and has a larger envelope due to varying the unknown coefficients $\{c_4, c_5, d_6\}$ for the double-squeezed and $\{c_4, c_5, d_3, d_4, d_5, d_6\}$ for the single-squeezed, in the range $[-100,100] \ {\rm Mpc}^2/{\rm h}^2$.
Note that in the single-squeezed case, $\epsilon=1$ corresponds to a degenerate configuration where all the vectors are in the same direction. Additionally, for our analysis, we choose $\vec{k}_2$ to be antiparallel to $\vec{k}_3$ with $|\vec{k}_2|=0.001$ h/Mpc.

The upper panels of Figs.~\ref{fig:ds} and~\ref{fig:ss} illustrate a number of interesting points. First, the contribution to the trispectrum from nonlinear clustering in the double-squeezed case can have significant dependence on $\theta$, as opposed to primordial contributions which are typically planar ($\theta=\pi$)~\cite{2015JCAP...05..007B}. Even though a complete study of primordial non-Gaussianity, including their effects on the LSS trispectrum, is beyond the scope of this work, we point out that features such as this angular dependence, could be useful to disentangle potential primordial signals from gravitational clustering effects.
Second, there are degeneracies and hierarchies among the counterterm contributions, and thus, in comparing to data with limited precision, it would be practical to reduce the operators to only the linear combinations that may be significantly detected. Said another way, the measurement of EFT coefficients may benefit from employing data with particular wavenumber configurations such that the signal from certain linear combinations of operators are enhanced.
The EFT predictions for the trispectrum in the lower panels of Figs.~\ref{fig:ds} and~\ref{fig:ss} have precision $\order(\Delta c \ k^2/k_{\rm NL}^2)$, where $\Delta c$ is the error in the EFT parameters. That is, to obtain the optimal precision of a one-loop calculation, {\it i.e.}~$\order(k^4/k_{\rm NL}^4)$,  the error in the EFT parameters propagated from the power spectrum and bispectrum must be further reduced. 

\section{Conclusion}
\label{sec:conclusions}
We have presented the first computation of the trispectrum in the EFT of LSS, highlighting important technical details that arise at this order.
First, we implemented the correct kernels for the vorticity and showed how vorticity arises from a convenient field redefinition. Second, we studied the impact of non-locality in time on the EFT counterterms and systematically showed that these can be parametrized by purely local operators. Third, we derived the stress tensor at $\mathcal{O}(k^2/k^2_\text{NL})$ and through NNLO in the linear density perturbations, and found the minimal basis of operators that is necessary for parametrizing any possible feedback between short and long scale modes. We applied all of these results to renormalize the trispectrum in a way that is consistent with the lower-order density correlators.  Lastly, we evaluated the trispectrum numerically for particular wavenumber configurations, giving a prediction for non-Gaussianity induced by non-linear structure formation.

There are a number of possible applications of our calculation of the trispectrum.  We have provided full SPT and EFT predictions at one-loop order for a quantity which has never been measured from LSS, and which has the potential to aid in probing the physics of inflationary non-Gaussianities. We have shown the form of both the SPT and EFT contributions in wavenumber configurations where inflationary non-Gaussianity is expected to be largest, thus effectively parametrizing the ``background'' of ordinary non-Gaussianities which come from gravitational mode-coupling. Furthermore, knowing the predicted analytic form of the trispectrum may facilitate the development of algorithms for measuring the trispectrum from N-body simulations and eventually LSS data. Finally, it may also be possible to measure some of the NNLO EFT coefficients by performing a full measurement of EFT coefficients present already in the two-loop power spectrum.

\vspace{0.1in}
\noindent {\large \em Acknowledgments}: 
We thank Simone Ferraro, Adrian Liu, Marcel Schmittfull, Uro\v{s} Seljak, Martin White, and Hojin Yoo for useful conversations pertaining to this work.  KS is supported by a Hertz Foundation Fellowship and by a National Science Foundation Graduate Research Fellowship. DB, MS, and KZ are supported under contract DE- AC02-05CH11231.

\appendix

\section{EFT Kernels}\label{app:kernels}
\label{EFTgraveyard}
Assuming the general form of the stress-tensor in Eq.~\eqref{eq:stresstensor}, the corresponding EFT kernels have been derived up to NNLO \cite{2015arXiv151207630B}. The $\widetilde{F}_1$ kernel can be written in terms of one independent operator, the $\widetilde{F}_2$ kernel in terms of three new independent operators in addition to the propagation of the one coming from the previous order, and finally the $\widetilde{F}_3$ kernel can be written in terms of eight new independent operators, in addition to the propagation of the four terms from the previous orders. The twelve operators can be chosen as those corresponding to $c_s^\delta$, $c_{1,2,3}^{\delta\delta}$, $c_{2,3}^{\theta\theta}$, and $c^{\delta\delta\delta}_{1,2,3,4,5,6}$ in Eq.~\eqref{eq:stresstensor}. We will collect here the expressions of $\widetilde{F}_1, \widetilde{F}_2$, and $\widetilde{F}_3$ in terms of this minimal set of operators.  

At LO we find
\begin{align} 
\widetilde{F}_1(\vec{k}) = -\frac{1}{9} c_s k^2,\quad\quad
\widetilde{G}_1(\vec{k}) = -\frac{1}{3}c_s k^2, 
\end{align}
where, for simplicity of notation, we have renamed $c_s^\delta=c_s$. At NLO we find

\begin{align*} \widetilde{F}_2(\vec{k}_1, \vec{k}_2) &= \,\frac{3}{11} \alpha (\vec{k}_1, \vec{k}_2) \left( \widetilde{G}_1 (\vec{k}_1) + \tilde{F}_1(\vec{k}_2)\right) + \frac{2}{33} \beta (\vec{k}_1, \vec{k}_2) \left(\widetilde{G}_1 (\vec{k}_1) + \widetilde{G}_1 (\vec{k}_2)\right)\\
&\quad -\frac{2}{33} \,c_s\left( k^2 F_2(\vec{k}_1, \vec{k}_2) -\vec{k}\cdot\vec{k}_2\right)- \frac{2}{33} \sum_{n=1}^3c_n\, k_ik_j\,e_n^{ij} (\vec{k}_1, \vec{k}_2),\\
\numberthis\label{eq:f2tilde} \\ 
\widetilde{G}_2(\vec{k}_1, \vec{k}_2)  &= \,\frac{1}{11} \alpha (\vec{k}_1, \vec{k}_2) \left( \widetilde{G}_1 (\vec{k}_1) + \widetilde{F}_1(\vec{k}_2)\right) + \frac{8}{33} \beta (\vec{k}_1, \vec{k}_2) \left(\widetilde{G}_1 (\vec{k}_1) + \widetilde{G}_1 (\vec{k}_2)\right)\\
&\quad -\frac{8}{33} \,c_s\left( k^2 F_2(\vec{k}_1, \vec{k}_2) -\vec{k}\cdot\vec{k}_2\right) - \frac{8}{33} \sum_{n=1}^3c_n\, k_ik_j\,e_n^{ij} (\vec{k}_1, \vec{k}_2),  
\numberthis\label{eq:g2tilde} \end{align*}
where $\vec{k}=\vec{k}_1+\vec{k}_2$, and $c_{1,2,3}$ are defined as in \cite{2015arXiv151207630B} to be
\begin{align}
\label{eq:c123}
c_1 = c_1^{\delta\delta}, \quad\quad
c_2 = c_2^{\delta\delta}+c_2^{\theta\theta}, \quad\quad
c_3 = c_3^{\delta\delta}+c_3^{\theta\theta}. \quad\quad
\end{align} 
Finally, at NNLO we find
\begin{align*}
\widetilde{F}_3(\vec{k}_1,\vec{k}_2,\vec{k}_3) &=  \, \frac{11}{52} \alpha( \vec{k}_1,  \vec{k}_2+ \vec{k}_3) \left[ \widetilde{G}_1 ( \vec{k}_1) F_2( \vec{k}_2,  \vec{k}_3) + \widetilde{F}_2 ( \vec{k}_2,  \vec{k}_3) \right] + \frac{11}{52} \alpha( \vec{k}_1+ \vec{k}_2,  \vec{k}_3) \Big[ \widetilde{G}_2( \vec{k}_1,  \vec{k}_2)\\
&\quad+G_2( \vec{k}_1,  \vec{k}_2) \widetilde{F}_1(\vec{k}_3)\Big] + \frac{1}{26} \beta( \vec{k}_1,  \vec{k}_2 +  \vec{k}_3) \left[\widetilde{G}_1 ( \vec{k}_1) G_2( \vec{k}_2,  \vec{k}_3) + \widetilde{G}_2 ( \vec{k}_2,  \vec{k}_3) \right]\\
&\quad + \frac{1}{26} \beta( \vec{k}_1+  \vec{k}_2 ,  \vec{k}_3) \left[ \widetilde{G}_2( \vec{k}_1,  \vec{k}_2) +G_2( \vec{k}_1,  \vec{k}_2) \widetilde{G}_1(\vec{k}_3)\right]\\
&\quad+\frac{1}{26} \beta^i_\omega (\vec{k}_1+\vec{k}_2, \vec{k}_3)\, \widetilde{G}_{2i}^{\omega}(\vec{k}_1, \vec{k}_2)
-\frac{11}{52} \alpha_\omega^i (\vec{k}_1 + \vec{k}_2, \vec{k}_3)\, \widetilde{G}_{2i}^{\omega} (\vec{k}_1, \vec{k}_2)\\
&\quad - \frac{1}{26 } c_s\left( k^2 F_3(\vec{k}_1, \vec{k}_2, \vec{k}_3)-\vec{k} \cdot (\vec{k}_2+\vec{k}_3) F_2(\vec{k}_2, \vec{k}_3)
+(1 - F_2(\vec{k}_1, \vec{k}_2) ) (\vec{k}\cdot\vec{k}_3)\right)\\
&\quad+\frac{1}{26} \sum_{n=1}^3 c_n\, k_i(k_{2}+k_{3})_j\,e_n^{ij} (\vec{k}_2, \vec{k}_3)-\frac{1}{26}  \sum_{n=1}^5c_n\, k_ik_j\,R_n^{ij} (\vec{k}_1,\vec{k}_2, \vec{k}_3)\\ 
&\quad-\frac{1}{26}  \sum_{n=1}^{6} d_n \,k_ik_j E_n^{ij} (\vec{k}_1, \vec{k}_2, \vec{k}_3),\numberthis \label{eq:f3tilde}
\end{align*}
where again $\vec{k}=\vec{k}_1+\vec{k}_2+\vec{k}_3$. The kernel $\widetilde{G}_{2i}^{\omega}$ gives the EFT NLO contribution to the vorticity (see Eq.~\eqref{eq:kernels}), and it is given by
\begin{align*}
&\widetilde{G}_{2i}^{\omega}(\vec{k}_1,\vec{k}_2)= -\frac{2}{9} \epsilon_{ijm} k^j\sum_{n=1}^3 c_n\,k_l\,e_n^{lm}(\vec{k}_1, \vec{k}_2). \numberthis\label{eq:Gomega}
\end{align*}
The functions $R_{1...5}^{ij}(\vec{k}_1,\vec{k}_2,\vec{k}_3)$ are defined as
\begin{align*}
R_1^{ij}(\vec{k}_1,\vec{k}_2,\vec{k}_3)&=F_2(\vec{k}_2,\vec{k}_3) \,e_1^{ij}(\vec{k}_1, \vec{k}_2+ \vec{k}_3)+F_2(\vec{k}_1,\vec{k}_2) \,e_1^{ij}(\vec{k}_1+\vec{k}_2, \vec{k}_3),\\
R_2^{ij}(\vec{k}_1,\vec{k}_2,\vec{k}_3)&=F_2(\vec{k}_2,\vec{k}_3) \,e_2^{ij}(\vec{k}_1, \vec{k}_2+ \vec{k}_3)+F_2(\vec{k}_1,\vec{k}_2) \,e_2^{ij}(\vec{k}_1+\vec{k}_2, \vec{k}_3),\\
R_3^{ij}(\vec{k}_1,\vec{k}_2,\vec{k}_3)&=\frac{5}{2}\left[F_2(\vec{k}_2,\vec{k}_3)-G_2(\vec{k}_2,\vec{k}_3)\right]e_2^{ij}(\vec{k}_1, \vec{k}_2+ \vec{k}_3)\\
&\quad+\frac{5}{2}\left[F_2(\vec{k}_1,\vec{k}_2)-G_2(\vec{k}_1,\vec{k}_2)\right]e_2^{ij}(\vec{k}_1+\vec{k}_2,\vec{k}_3)\\
&\quad-\frac{1}{2}\left[3F_2(\vec{k}_2,\vec{k}_3)-5G_2(\vec{k}_2,\vec{k}_3)\right]e_3^{ij}(\vec{k}_1, \vec{k}_2+ \vec{k}_3)\\
&\quad-\frac{1}{2}\left[3F_2(\vec{k}_1,\vec{k}_2)-5G_2(\vec{k}_1,\vec{k}_2)\right]e_3^{ij}(\vec{k}_1+\vec{k}_2,\vec{k}_3),\\
R_4^{ij}(\vec{k}_1,\vec{k}_2,\vec{k}_3)&=\left[G_2(\vec{k}_2,\vec{k}_3)-F_2(\vec{k}_2,\vec{k}_3)\right]e_2^{ij}(\vec{k}_1,\vec{k}_2+\vec{k}_3)\\
&\quad+\left[G_2(\vec{k}_1,\vec{k}_2)-F_2(\vec{k}_1,\vec{k}_2)\right]e_2^{ij}(\vec{k}_1+\vec{k}_2,\vec{k}_3),\\
R_5^{ij}(\vec{k}_1,\vec{k}_2,\vec{k}_3)&=\left[G_2(\vec{k}_2,\vec{k}_3)-F_2(\vec{k}_2,\vec{k}_3)\right]e_3^{ij}(\vec{k}_1,\vec{k}_2+\vec{k}_3)\\
&\quad+\left[G_2(\vec{k}_1,\vec{k}_2)-F_2(\vec{k}_1,\vec{k}_2)\right]e_3^{ij}(\vec{k}_1+\vec{k}_2,\vec{k}_3).\\
\end{align*}
The coefficients $c_{1,2,3}$ are defined in Eq.~\eqref{eq:c123}, and $c_{4,5}$ are defined as in \cite{2015arXiv151207630B} to be
\begin{align}
\label{eq:c45}
c_4 = c_2^{\theta\theta}+\frac{5}{2}(c_3^{\delta\delta}+c_3^{\theta\theta}), \quad\quad
c_5 = c_3^{\theta\theta}-\frac{5}{2}(c_3^{\delta\delta}+c_3^{\theta\theta}),
\end{align} 
and for simplicity of notation we have renamed $c^{\delta\delta\delta}_{1...6}=d_{1...6}$.
For completeness, in Eqs.~\eqref{eq:f2tilde}, \eqref{eq:g2tilde}, and \eqref{eq:f3tilde} we have included the expansion of the $(1 + \delta)^{-1}$ term which appears in the equations of motion. Note that the kernels in Eqs.~(\ref{eq:f2tilde}-\ref{eq:Gomega}) are not symmetric in their arguments, and need to be symmetrized when used to calculate amplitudes.

\bibliographystyle{JHEP}
\bibliography{Trispectrum}
\end{document}